\documentclass[11pt]{article}
\pdfoutput=1
\usepackage[margin=1.0in]{geometry}
\usepackage{setspace}
\usepackage{amsmath,amssymb,graphicx}
\usepackage[pagebackref=true]{hyperref}
\usepackage{slashed}

\usepackage[T1]{fontenc}

\usepackage[utf8]{inputenc}

\usepackage[greek,english]{babel}

\usepackage[font=small,labelfont=bf]{caption}
\usepackage{cite}

\DeclareMathOperator{\tr}{tr}

\newcommand{\iu}{{\mathrm i}}
\newcommand{\E}{{\mathrm e}}
\newcommand{\cpi}{\text{\greektext p}}

\newcommand{\rmd}{\mathrm{d}}
\newcommand{\dif}[1]{\ensuremath{\operatorname{d}\!{#1}}}

\newcommand{\Uone}{{\mathrm{U(1)}}}
\newcommand{\SU}{{\mathrm{SU}}}
\newcommand{\SUN}{{\mathrm{SU}(N)}}

\newcommand{\cT}{{\cal T}}

\newcommand{\be}{\begin{equation}}
\newcommand{\ee}{\end{equation}}

\newcommand{\ZZ}{{\mathbb{Z}}}

\newcommand{\Cp}{{C^{(p)}}}

\usepackage{xcolor}
\usepackage{bm}

\usepackage{tikz}
\usetikzlibrary{calc}

\date{April 24, 2025} 
\title{\bf 
Extra-Dimensional Axion Expectations}
\author{Matthew Reece \\
{\small \color{gray} \texttt{mreece~(@g.harvard.edu)}}\\
{\small Department of Physics, Harvard University, Cambridge, MA, 02138}}

\begin{document}
\maketitle

\begin{abstract}
Axions arising as modes of higher-dimensional gauge fields are known to offer a compelling solution to the axion quality problem and to naturally arise in string theory. In this context, it is interesting to ask how we would interpret an experimental measurement of the axion decay constant $f$. I give several arguments for, as well as concrete examples in string theory of, the existence in such a model of an axion string with tension of order $2\cpi S_\mathrm{inst} f^2$, where $S_\mathrm{inst}$ is the instanton action. Furthermore, in models of this type axion strings are typically fundamental objects (rather than solitons), whose tension is at or above the fundamental cutoff of the theory. As a result, I argue that for an extra-dimensional QCD axion, it is likely that the fundamental cutoff scale lies at most two orders of magnitude above $f$. In addition to these core arguments, this paper begins with a self-contained introduction to the physics of extra-dimensional axions and ends with some comments on axion physics in relation to chiral fermions.
\end{abstract}

\tableofcontents

\section{Introduction and Summary}

The most well-studied solution to the Strong CP problem is an axion field, namely a field $\theta(x)$ of period $2\cpi$ that couples to the gluon field strength $G$ via the interaction
\begin{equation} \label{eq:axiongluon}
S_\mathrm{axion} = \frac{k_G}{8\cpi^2} \int \theta(x) \tr\left[G(x) \wedge G(x)\right],
\end{equation}
for some integer $k_G$. If this is the only interaction that violates the continuous shift symmetry $\theta \mapsto \theta + \text{const.}$, then strong dynamics generates an effective potential that dynamically relaxes the QCD theta angle to zero~\cite{Peccei:1977ur, Peccei:1977hh, Wilczek:1977pj, Weinberg:1977ma, DiVecchia:1980yfw, Vafa:1984xg}. The original axion model involved spontaneous breaking of an approximate U(1) global symmetry at the electroweak scale, and was soon ruled out by data. Shortly thereafter, three distinct axion scenarios with much weaker interactions, safe from experimental limits at the time, were proposed. The first two, the KSVZ~\cite{Kim:1979if, Shifman:1979if} and DFSZ~\cite{Zhitnitsky:1980tq, Dine:1981rt} models, posit the axion as a pseudo-Nambu-Goldstone boson (PNGB) of an approximate U(1) (``Peccei-Quinn'' or PQ) global  symmetry that is spontaneously broken at a high scale. They differ in whether or not the Standard Model Higgs and fermion fields are charged under this symmetry (no for KSVZ, yes for DFSZ). These are the best-known paradigms for axion models. They are widely taught, and minimal incarnations of each have become standard benchmarks on experimental sensitivity plots. (The basic KSVZ and DFSZ paradigms can be realized with a variety of choices of field content, predicting different values for the axion--photon coupling, so one should not take the standard plot labels too seriously.)

The third axion scenario, the extra-dimensional axion~\cite{Witten:1984dg}, is the subject of this paper. It was proposed just a few years after the KSVZ and DFSZ scenarios, and has very different physics. There is no spontaneous breaking of a 4d U(1) symmetry to produce a pseudo-Nambu-Goldstone boson. Instead, the axion is a mode of a higher-dimensional gauge field. These models are well-motivated, from the bottom-up perspective (they significantly ameliorate the axion quality problem), from the top-down perspective (the ingredients automatically arise in string theory), and from more general principles (such axions appear to play a key role for the consistency of quantum gravity). Despite the large literature on such models, they are under-appreciated, particularly in discussions of axion phenomenology and cosmology. For example, because these models do not have a 4d PQ-breaking phase transition, the often-discussed post-inflationary axion cosmology scenario does not apply to them. Furthermore, the physics at the scale of the axion decay constant $f$ is dramatically different: rather than simply expecting to find some new 4d fields with masses of order $f$, one expects that $f$ is near the quantum gravity cutoff scale, where local quantum field theory breaks down entirely. If an axion particle is discovered and its decay constant is measured, then, the conclusions we would draw from this measurement are radically different in the extra-dimensional scenario and in the more widely-studied 4d scenarios. 

I have two main goals in this paper. First, I give a general exposition of the extra-dimensional axion scenario, based on its core ingredients rather than starting from a specific string theory realization. Much of this exposition covers familiar territory, though there are certain details (e.g., the argument for the existence of massless axions in a compactification with general warping) that I have not seen spelled out explicitly before (though they are surely known to experts). Second, I explain some expectations for how various energy scales are related to each other in extra-dimensional axion models. In particular, I emphasize that in the extra-dimensional scenario (unlike the KSVZ or DFSZ scenarios) the axion decay constant is likely to carry information about the fundamental scale where quantum field theory breaks down. This raises the stakes for an experimental discovery of an axion, but also points to the need for observational signals that could convincingly distinguish between a 4d axion and an extra-dimensional one.

Before summarizing the claims of this paper in more detail, I offer a brief (necessarily incomplete) review of the literature on extra-dimensional axions. To the best of my knowledge, the extra-dimensional axion was first discussed in the context of Type I string theory~\cite{Witten:1984dg}, where the necessary ingredients appeared automatically, rather than being constructed specifically to solve the Strong CP problem. Such axions received significant attention in studies of Type I and heterotic string theory~\cite{Choi:1985je, Barr:1985hk, Choi:1985bz, Dine:1986bg, Dine:1986zy, Dine:1987bq, Banks:1996ss, Banks:1996ea}. The potential for string-theoretic axions to ameliorate the axion quality problem was emphasized in~\cite{Kallosh:1995hi}. Later, it was understood that constructions of gauge fields on D-branes in Type II string theory also automatically contain axion-like couplings to RR gauge fields~\cite{Douglas:1995bn, Vafa:1995bm, Green:1996dd}. Following the initial wave of research on moduli stabilization in string theory, two key papers made strides toward bridging the gap between string theory constructions and axion phenomenology~\cite{Conlon:2006tq, Svrcek:2006yi}. A few papers introduced the idea of extra-dimensional axions into bottom-up phenomenological model-building~\cite{Cheng:2001ys, Arkani-Hamed:2003xts, Choi:2003wr}. Eventually, the fact that string theory can predict a significant number of axion-like fields (already observed in~\cite{Witten:1984dg}) became widespread knowledge, with phenomenological studies emphasizing that the axion decay constant in such a scenario was likely to be large (with associated cosmological difficulties~\cite{Fox:2004kb}) and that the number of axion-like particles could be large~\cite{Arvanitaki:2009fg}. Phenomenological interest, in turn, fed back into string theory studies~\cite{Acharya:2010zx, Cicoli:2012sz}, which have recently made computational advances allowing the exploration of phenomenological implications of a large number of axion fields (e.g.,~\cite{Demirtas:2018akl, Halverson:2019cmy, Mehta:2021pwf, Broeckel:2021dpz, Demirtas:2021gsq, Gendler:2023kjt}). Recent work has emphasized that although 4d axion models produce solitonic axion strings, with PQ symmetry restored in their core, by contrast axion strings for extra-dimensional axions are not solitons and have cores that probe deep ultraviolet physics and infinite-distance limits of moduli space~\cite{Dolan:2017vmn, Reece:2018zvv, Lanza:2020qmt, Lanza:2021udy, March-Russell:2021zfq}. This leads to qualitative cosmological differences; in particular, there is no PQ-violating phase transition in which axion strings would be produced by the Kibble-Zurek mechanism~\cite{Cicoli:2022fzy, Benabou:2023npn}. Other recent work discussing differences between extra-dimensional and 4d axion models includes~\cite{Choi:2021kuy, Adachi:2021rjw, Burgess:2023ifd, Reece:2023czb, Choi:2024ome}. A recent talk by Kiwoon Choi~\cite{Choi:2023talk} took a similar perspective to my own, proposing the terminology ``Type I axion'' and ``Type II axion'' for 4d and extra-dimensional axions respectively and emphasizing their differences.

\medskip 

Here is a brief summary of the main messages to take away from the paper; some review well-known facts, while others are novel:

\begin{itemize}

\item {\bf \em Extra-dimensional gauge theories automatically produce high-quality axions.} Specifically, given a U(1) $p$-form (antisymmetric $p$-index tensor) gauge field $C_{\mu_1 \cdots \mu_p}$, there is a 4d axion mode, schematically of the form $\int_{\Sigma} C_{\mu_1 \cdots \mu_p} \dif{x^{\mu_1}}\cdots \dif{x^{\mu_p}}$, for each independent, non-torsion, $p$-dimensional cycle $\Sigma$ contained in the extra dimensions. This is true in complete generality, including in warped compactifications. The 4d axion is a periodic scalar field that obtains a mass only through exponentially small instanton effects, corresponding either to traditional 4d gauge theory instantons or to semiclassical effects from wrapping extended objects around the extra dimensions. 

\item {\bf \em There is no Peccei-Quinn phase transition for an extra-dimensional axion.} In these models, the axion is not a PNGB of an anomalous 4d symmetry, except in the trivial sense that the axion particle can be created by the shift-symmetry current $\partial_\mu \theta(x)$. In particular, the analogue of the PQ symmetry is already present in the higher dimensional theory and acts {\em nonlinearly} on the gauge field, descending to an approximate shift symmetry on $\theta$ in 4d. There is no analogue of the usual process by which a complex scalar gets a VEV as the universe cools, breaking PQ symmetry. Importantly, this means that {\bf \em extra-dimensional axions intrinsically belong to the ``pre-inflationary axion'' class of cosmologies}---the PQ symmetry is always already broken---and the most serious cosmological problem with such models is the axion isocurvature problem rather than the domain wall problem.

\item {\bf \em The axion decay constant is a guide to the fundamental UV cutoff of the theory.} This claim, in quantitative form, is the main novel content of the paper. I will give several (non-rigorous) arguments and survey several examples in string theory in support of the claim that, given an axion coupled to gauge fields with associated instanton action $S_\mathrm{inst}$ and an axion decay constant $f$, we expect to find a (possibly nonminimal) axion string with core tension near the scale
\begin{equation} \label{eq:axiontensionestimate1}
\cT \approx 2\cpi S_\mathrm{inst} f^2.
\end{equation}
In theories where the instanton expansion is under control, and particularly for a QCD axion where $S_\mathrm{inst} = 8\cpi^2/g^2$ (for a UV value of the QCD coupling $g$), this is significantly larger than the naive 4d field theory estimate $\cT \sim \cpi f^2$. It is also, in general, significantly {\em lower} than the upper bound $\cT \sim 2\cpi f M_\mathrm{Pl}$ allowed by the magnetic axion Weak Gravity Conjecture (WGC)~\cite{Arkani-Hamed:2006emk}.

This bound is important because, in theories of an extra-dimensional axion, we expect that the axion strings are fundamental objects (e.g., fundamental strings, or D-branes wrapped on internal dimensions), and that their tension scale lies at or above the quantum gravity cutoff $\Lambda_\textsc{QG}$ at which local quantum field theory completely breaks down. If an axion is discovered and we measured $f$, then, we can estimate that---{\em assuming an extra-dimensional axion}---$\Lambda_\textsc{QG}$ is no larger than $\sim 10^2 f$. This is a powerful claim, as currently the best that we can say from general principles is that $\Lambda_\textsc{QG} \lesssim M_\mathrm{Pl}/\sqrt{N_\textsc{SM}}$ where $N_\textsc{SM}$ is the number of weakly-coupled fields in the Standard Model~\cite{Veneziano:2001ah}.

\item {\bf \em In theories with charged chiral matter, like the Standard Model, it is difficult to turn on tree-level mass terms that entirely remove a light axion from the spectrum.} There are two types of tree-level mass that an axion can have: a monodromy mass of the form $\theta F^{(4)}$, and a Stueckelberg mass of the form $|A - \rmd \theta|^2$. We argue, in the context of a particular string realization, that the former clashes with the existence of chiral matter charged under gauge fields the axion couples to. We also argue that the latter case typically involves not only an extra-dimensional axion, but a second axion arising from the phase of a charged matter field. This scenario can interpolate between an extra-dimensional axion and a conventional 4d axion, depending on which linear combination of the two is eaten and which remains light. These considerations support the robustness of the existence of a light axion interacting with the Standard Model gauge fields.

\end{itemize}

The remainder of the paper is organized as follows. In \S\ref{sec:xdaxions}, I provide a self-contained exposition of the general theory of axions arising as zero modes of $p$-form gauge fields in higher dimensions, as well as the origin of axion-gauge field couplings, instantons, and axion strings in this setting. The contents of this section are mostly review of well-known physics. The next three sections constitute the main original content of the paper. In \S\ref{sec:uvcutoffestimate}, I give a simple effective field theory estimate of an upper bound on the UV cutoff of a theory of an extra-dimensional axion, scaling like the ratio $f/g$ of the axion decay constant to the gauge theory coupling. In \S\ref{sec:arguments}, I provide arguments for the estimate~\eqref{eq:axiontensionestimate1} for the axion string tension. In \S\ref{sec:examples}, I give explicit examples of axion string tensions computed in string theory that substantiate this estimate, finding an interesting twist in one case where the relevant axion string is nonminimal. In \S\ref{sec:chiralfermions}, I offer a few comments on the relationship between axion physics and the existence of chiral fermions in the context of extra-dimensional axions. In particular, I argue that tree-level mass terms are unlikely to remove light axions from the spectrum. I also review an argument that, for extra-dimensional axions, the derivative couplings of axions to Standard Model fermions are likely to be of one-loop size, in contrast to DFSZ models (where they appear at tree level) and KSVZ models (where they appear at two loops). In \S\ref{sec:outlook}, I conclude with a list of questions and future directions to consider.

\section{Axions from higher-dimensional gauge fields}
\label{sec:xdaxions}
\subsection{What is an extra-dimensional axion?}
\label{subsec:whatisxdaxion}

The basic idea that a zero mode of a higher-dimensional gauge field can give rise to an axion is likely familiar to many readers. For instance, given an ordinary $\Uone$ gauge field in a 5d theory compactified on a circular direction $y \cong y + 2\cpi R$, the field $\theta(x) \equiv \int_0^{2\cpi R} A_y(x,y) \dif y$ behaves as a periodic scalar $\theta \cong \theta + 2\cpi$. In this section, I will derive the general case:

\medskip
{\noindent \bf \em Existence of extra-dimensional axions.} Suppose that spacetime has $d = (4+n)$ dimensions in the form of a warped product compactification, $M = X \times_w Y$ for a 4-dimensional spacetime $X$ and an $n$-dimensional space $Y$. If there is a $p$-form gauge field $C^{(p)}$ on $M$, there is a distinct massless, periodic 4d axion field for every independent non-torsion $p$-cycle in $Y$.
\medskip

\noindent
Let's explain this step by step, with a very quick review of several relevant concepts in mathematics and physics. Readers who find this difficult to follow might want to first review portions of my TASI lecture notes~\cite{Reece:2023czb}.

\subsubsection{Geometry and topology}

Throughout this paper, all manifolds are assumed to be oriented. Physically, this is equivalent to saying that we assume that our theories do not have an orientation-reversing spacetime symmetry, like parity or CP. This appears to be true of the world that we live in, at least at low energies, and mathematically it underlies many assumptions below. It would be interesting to revisit all of these results in the non-orientable case. (See~\cite{McNamara:2022lrw} for a recent introduction to physics on non-orientable manifolds, aimed at particle physics applications.) We will also assume that our manifolds are spin, since we are interested in real-world theories admitting fermions.

A warped compactification $M = X \times_w Y$ is the most general compactification preserving the symmetries of $X$ (e.g., Poincar\'e symmetry, if we take $X$ to be 4d Minkowski spacetime). It has a metric 
\begin{equation}
\dif s^2 = w(y) \dif s^2_X + \dif s^2_Y,
\end{equation}
where $w(y) \geq 0$. The amount of warping is measured by $\dif w(y)$. A useful geometric fact is that the Hodge star on a warped product manifold is a warped product of Hodge stars, i.e., if $\eta^{(q)}_X$ is a $q$-form on $X$ and $\xi_Y^{(r)}$ is an $r$-form on $Y$ then
\begin{align}\label{eq:hodgeproduct}
\star\left(\eta^{(q)}_X \wedge \xi^{(r)}_Y\right) = (-1)^{q r} w(y)^{\mathrm{dim}(X)/2 - q} \left(\star_X \eta^{(q)}_X \wedge \star_Y \xi^{(r)}_Y\right).
\end{align}
If we perform a Weyl rescaling of the metric on $Y$, the Hodge star acting on an $r$-form changes as follows:
\begin{equation} \label{eq:weylhodge}
\widetilde{\dif s^2_Y} \equiv f(y) \dif s^2_Y \quad \Rightarrow \quad \widetilde{\star_Y} \xi^{(r)}_Y = f(y)^{n/2 - r} {\star_Y \xi^{(r)}_Y}.
\end{equation}
The factor of $f(y)^{n/2}$ comes from $\sqrt{|\det \widetilde{g_Y}|}$ and the factor $f(y)^{-r}$ from $r$ factors of $\widetilde{g_Y}^{\mu \nu}$.

The topology of $Y$ can be characterized by homology classes (cycles) in $H_p(Y, \mathbb{Z})$. These cycles correspond to integer linear combinations of equivalence classes of $p$-dimensional submanifolds without boundary, modulo boundaries. Linear combinations of spaces may seem counterintuitive, so one can also think of these classes in terms of {\em integrals} over those spaces, with the equivalence $\int_{\Sigma^{(p)}} \sim \int_{\Sigma^{(p)}} + \int_{\partial \Lambda^{(p+1)}}$ holding (due to Stokes's theorem) whenever the form that we integrate is closed. We choose a basis of $p$-cycles $[\Sigma^{(p)}_i]$ for the torsion-free part of $H_p(Y, \ZZ)$, where torsion-free means that there is no integer $n \neq 0$ such that $n[\Sigma^{(p)}] = 0$. The number of independent torsion-free cycles is also known as the $p^\textrm{th}$ Betti number $b_p(Y)$. There are corresponding (non-torsion) cohomology classes $[\omega^{(p)}_i] \in H^p(Y, \ZZ)$, which are equivalence classes of closed differential forms  ($\dif \omega = 0$) which give integers when integrated over cycles in $H_p(Y, \ZZ)$, modulo the addition of exact forms ($\omega \sim \omega + \dif \lambda$). We can choose our bases in homology and cohomology to be related, such that
 \begin{equation} \label{eq:pformbasis}
 \int_{\Sigma^{(p)}_i} \omega^{(p)}_j = \delta^i_j.
 \end{equation}
This integral is independent of the choice of differential form $\omega^{(p)}_j$ from the cohomology class $[\omega^{(p)}_j]$ (because $\partial \Sigma^{(p)}_i = 0$) and of the choice of submanifold $\Sigma^{(p)}_i$ from the homology class $[\Sigma^{(p)}_i]$ (because $\dif \omega^{(p)}_j = 0$). 

Each cohomology class contains a unique harmonic representative, i.e., there is precisely one $\omega_p \in [\omega_p]$ such that
\begin{equation}
{\dif \star_Y \omega_p} = 0.
\end{equation}
(A form is harmonic if and only if $\dif \omega = 0$ {\em and} $\dif \star_Y \omega = 0$.) Notice that the definition of harmonic depends on a choice of metric, via $\star_Y$.

\subsubsection{$p$-form gauge theory on $M$}

We'll consider a theory with a $p$-form U(1) gauge field
 \begin{equation}
 C^{(p)} \equiv \frac{1}{p!} C_{\mu_1 \cdots \mu_p} \dif x^{\mu_1} \wedge  \cdots \wedge \dif x^{\mu_p}.
 \end{equation}
As usual, when we introduce a gauge field, we require invariance under local gauge transformations $C^{(p)} \mapsto C^{(p)} + \dif \lambda^{(p-1)}$ for any $(p-1)$-form $\lambda$. However, we also have to allow for the case where $\lambda^{(p-1)}$ is not a single-valued form, i.e., when it has some sort of ``winding'' around extra dimensions. We can make this more precise using the cohomology classes introduced above. First, it is useful to introduce Wilson loops, as a probe for the right notion of gauge invariance in the theory.
 
Associated to the gauge field is a family of Wilson-loop operators: given a U(1) charge $q \in \ZZ$ and a $p$-dimensional submanifold $\Xi^{(p)}$, we have an operator
 \begin{equation} \label{eq:wilsonloop}
 W_q(\Xi^{(p)}) \equiv \exp\left[\iu q \int_{\Xi^{(p)}} C^{(p)}\right].
 \end{equation}
The Wilson loop operator is not topological; it depends on the actual geometry of $\Xi^{(p)}$. (This is familiar in 4d gauge theory, where we commonly discuss perimeter or area laws for Wilson loops.) Nonetheless, it is useful to decompose the homology class of $\Xi^{(p)}$ in our chosen basis, $[\Xi^{(p)}] = \sum_j m_j \Sigma_j^{(p)}, m_j \in \ZZ$.
 Gauge transformations of $C^{(p)}$  preserve all of the Wilson loops. These include the local gauge transformations $C^{(p)} \mapsto C^{(p)} + \dif \lambda^{(p-1)}$. However, they also include the ``large'' gauge transformations,
 \begin{equation} \label{eq:largegauge}
 C^{(p)} \mapsto C^{(p)} + 2\cpi \sum_{i=1}^{b_p(Y)} n^i \omega^{(p)}_i, \quad n^i \in \ZZ.
 \end{equation}
Under this operation, we have
\begin{align}
 W_q(\Xi^{(p)}) &\mapsto W_q(\Xi^{(p)})\exp\left[\iu q \sum_j \left(m_j \int_{\Sigma^{(p)}_j} 2\cpi \sum_i n^i \omega^{(p)}_i\right)\right]
 \nonumber \\ 
 & =  W_q(\Xi^{(p)}) \exp\left[2\cpi \iu q \sum_j m_j n^j\right] = W_q(\Xi^{(p)}), 
\end{align}
using~\eqref{eq:pformbasis}. These large gauge transformations are the higher-form analogue of familiar winding gauge transformations in the case of ordinary 1-form U(1) gauge fields. They are important for our story because they descend to periodicity conditions on the 4d axions.

Readers for whom these ``large'' gauge transformations are unfamiliar are recommended to think through the case of a 5d compactification on a circle $S^1$ parametrized by a coordinate $y \cong y + 2\cpi R$, where the homology class $\Sigma^{(1)}$ corresponding to the circle is dual to the class $[\dif y/(2\cpi R)] \in H^1(S^1, \ZZ)$ in cohomology. Despite the notation, $\dif y$ is {\em not} exact, because $y$ is not a well-defined (single-valued) function on the circle. This is the prototypical example of a U(1) gauge transformation with winding.
 
\subsubsection{Existence of massless 4d axion modes}

We consider the standard kinetic term for our $p$-form gauge field in $d = 4+n$ dimensions,
\begin{equation} \label{eq:kineticd}
-\int_M \frac{1}{2 e_p^2} \Phi(x,y) \dif \Cp(x,y) \wedge \star \dif \Cp(x,y).
\end{equation}
where $e_p$ is the $p$-form gauge coupling and $\Phi(x,y)$ is a scalar modulus field. We assume $\Phi(x,y) \geq 0$ to avoid instabilities arising from wrong-sign kinetic terms. In our conventions, $C_{\mu_1 \cdots \mu_p}$ has mass dimension $p$, so $\Cp$ is dimensionless, $\star \dif \Cp$ has dimension $2p - d + 1$, and $e_p$ has mass dimension $p + 1 - \frac{d}{2}$. The equation of motion of $\Cp$, in the limit that only the kinetic term matters, is simply
\begin{equation} \label{eq:higherdeom}
\dif{\left(\Phi{\star}{\dif \Cp}\right)} = 0.
\end{equation}
Furthermore, we assume that there is a background solution in which $\Phi = \Phi(y)$ is independent of $x$. 

Now, we consider an ansatz for a 4d axion field perturbing around this background. Specifically, we consider
\begin{equation}
\Cp(x,y) = \theta(x) \omega^{(p)}(y),
\end{equation}
where $\theta$ is a scalar function on $X$ and $\omega^{(p)}$ is a $p$-form on $Y$. Then
\begin{equation} \label{eq:dCp}
\dif\Cp(x,y) = \dif \theta(x) \wedge \omega^{(p)}(y) + \theta(x) \dif\omega^{(p)}(y).
\end{equation}
We are interested in searching for massless axion fields, for which the kinetic term~\eqref{eq:kineticd} should depend only on $\dif \theta(x)$ and not $\theta(x)$. We see that this motivates searching for solutions where $\dif\omega^{(p)} = 0$, i.e., where $\omega^{(p)}$ is a closed $p$-form on $Y$, so that we can drop the second term in~\eqref{eq:dCp}.

Next, we require that our ansatz obeys the equation of motion~\eqref{eq:higherdeom} for the higher-dimensional theory. We have
\begin{align}
\dif{\left(\Phi{\star}{\dif \Cp}\right)} &= \dif{\left[ \Phi(y){\star}{\left(\dif{\theta(x)}\wedge\omega^{(p)}(y)\right)}\right]} = \dif{\left[{\star}\left(\dif{\theta}\wedge(\Phi(y) \omega^{(p)}(y))\right)\right]} \nonumber \\
&= (-1)^p \dif{\left[w(y) {\star_X}{\dif{\theta}} \wedge {\star_Y}{(\Phi(y) \omega^{(p)}(y))}\right]} \nonumber \\
&= (-1)^p  \left\{w(y) \Phi(y) (\dif{\star_X}{\dif \theta}) \wedge {\star_Y}{\omega^{(p)}} - {\star_X}{\dif{\theta}} \wedge \dif{\left[w(y) \Phi(y) {\star_Y{\omega^{(p)}}}\right]}\right\} = 0.
\end{align}
In moving from the first line to the second we used~\eqref{eq:hodgeproduct}. In order for the final line to match the expected 4d equation of motion of a massless scalar field, $\dif{\star_X}{\dif \theta} = 0$, we should require that
\begin{equation} \label{eq:warpedharmonic}
\dif{\left[w(y) \Phi(y) {\star_Y{\omega^{(p)}}}\right]} = 0.
\end{equation}
In the case $w(y) = \Phi(y) = 1$, this is just the condition that $\omega^{(p)}$ be a {\em harmonic form}. (More precisely, it's the condition that $\omega^{(p)}$ is co-closed, but because $\omega^{(p)}$ was already assumed to be closed, this is equivalent to being harmonic.) As noted above, there is always a unique choice of such a form in every cohomology class. But suppose we have nontrivial warping $w(y)$ or a nontrivial modulus profile $\Phi(y)$ in the extra dimensions. It is still true that a unique solution to~\eqref{eq:warpedharmonic} exists for a given cohomology class; we will call such a representative a warped harmonic form. (Here and below I will neglect $\Phi(y)$ in my terminology by saying ``warped'' instead of ``warped and modulated,'' for brevity.) 

One way to show that a unique warped harmonic form exists is to simply repeat the usual argument in the unwarped case (e.g.,~\cite{rosenberg1997laplacian} for details, or~\cite[\S7.9]{Nakahara:2003nw} for a summary) with a minor modification. First, we define a symmetric, positive definite warped inner product on $r$-forms on $Y$ by 
\begin{equation} \label{eq:warpedinner}
(\xi, \eta) = \int_Y w(y)\Phi(y)\, \xi(y) \wedge \star_Y \eta(y),
\end{equation}
that is, we weight the usual definition by the (positive) factor $\Phi(y) w(y)$. Then we can define an adjoint exterior derivative operator taking $r$-forms on $Y$ to $(r-1)$-forms on $Y$, via 
\begin{equation}
\mathrm{d}^\dagger\xi^{(r)} \equiv (-1)^{n(r+1)} \frac{1}{w(y)\Phi(y)} \star_Y \dif{\left(w(y) \Phi(y) {\star_Y}\xi^{(r)}\right)}.
\end{equation}
This is an adjoint with respect to our inner product, in the sense that $(\dif \xi, \eta) = (\xi, \mathrm{d}^\dagger\eta)$ for an $(r-1)$-form $\xi$ and $r$-form $\eta$. We then define the warped Laplacian $\Delta = \mathrm{d}^\dagger \mathrm{d} + \mathrm{d} \mathrm{d}^\dagger$. The usual Hodge decomposition theorem then follows by precisely the usual logic, with our modified definitions. A warped harmonic form with $\Delta \omega = 0$ is closed and warped co-closed, i.e., $\mathrm{d}^\dagger \omega = 0$, which means it obeys the condition~\eqref{eq:warpedharmonic}. 

If $n \neq 2p$, we can give a simple alternative argument. Define an auxiliary metric, which is a Weyl rescaling of our original metric:
\begin{equation}\label{eq:auxmetric}
\widehat{\dif s_Y}^2 \equiv \left[w(y) \Phi(y)\right]^{2/(n-2p)} \dif s_Y^2.
\end{equation}
By~\eqref{eq:weylhodge}, we see that the condition~\eqref{eq:warpedharmonic} is the same as the condition
\begin{equation}\label{eq:auxharmonic}
\dif{\left(\widehat{\star_Y} \omega^{(p)}\right)} = 0.
\end{equation}
In other words, we can find the desired form $\omega^{(p)}$ for our extra-dimensional ansatz by choosing the unique representative in its cohomology class that is harmonic with respect to the auxiliary metric~\eqref{eq:auxmetric} rather than the original metric on $Y$.

\subsubsection{Axion kinetic matrix and periodicity}

Now that we have seen how to write down an ansatz for a $d$-dimensional $\Cp$ field that leads to a 4-dimensional axion $\theta$, let's consider the general case where we have a collection of such axions. Given a basis of cohomology classes for $H^p(Y,\ZZ)$ as in~\eqref{eq:pformbasis}, let $\widehat\omega^{(p)}_i(y) \in [\omega^{(p)}_i]$ be the unique representative in its class that is a warped harmonic form as in~\eqref{eq:warpedharmonic}. Then we consider the ansatz
\begin{equation} \label{eq:Cpansatz}
\Cp(x,y) = \sum_{i=1}^{b_p(Y)} \theta^i(x) \widehat\omega^{(p)}_i(y).
\end{equation}
Substituting this into the kinetic term~\eqref{eq:kineticd}, together with the choice $\Phi = \Phi(y)$ (independent of $x$), we obtain the four-dimensional kinetic term
\begin{align} \label{eq:axionkinetic}
S_\mathrm{kin} &= -\int_X \frac{1}{2} \kappa_{ij} \dif \theta^i(x) \wedge \star_X \dif \theta^j(x), \qquad\text{where} \nonumber \\
\kappa_{ij} &\equiv \frac{1}{e_p^2} \left(\widehat\omega^{(p)}_i, \widehat\omega^{(p)}_j\right) = \frac{1}{e_p^2} \int_Y w(y)\Phi(y) \,\widehat\omega^{(p)}_i \star_Y \widehat\omega^{(p)}_j.
\end{align}
In $S_\mathrm{kin}$ we have left the sum over $i, j \in \{1, \ldots, b_p(Y)\}$ implicit. The warping $w(y)$ and the modulus $\Phi(y)$ appear only through the definition of the warped inner product~\eqref{eq:warpedinner} and the choice of the warped harmonic representatives $\widehat\omega^{(p)}_i$. The matrix $\kappa_{ij}$ has mass dimension $2$.

The redundancy~\eqref{eq:largegauge} under ``large'' gauge transformations becomes precisely a periodicity constraint on the 4d axions:
\begin{equation} \label{eq:axionperiod}
\theta^i(x) \cong \theta^i(x) + 2\cpi n^i, \quad n^i \in \ZZ.
\end{equation}
In other words, our conventions are such that every axion field is $2\cpi$-periodic; we should think of~\eqref{eq:axionperiod} as a gauge transformation on the axion fields. This is the only remnant of the $p$-form gauge invariance in the 4d axion theory. In particular, the Wilson loop operator $W_q(\Sigma^{(p)}_i)$ matches to the 4d local operator $\exp(\iu q \theta^i)$, which is gauge invariant. We can interpret the diagonal entries of the kinetic matrix as squared decay constants, $f_{i} \equiv \sqrt{\kappa_{ii}}$, such that the period of the canonically normalized axion field is $2\cpi f_i$. The off-diagonal entries are kinetic mixing parameters~\cite{Babu:1994id}.

Finally, we note that we can also think of the individual 4d axion fields in terms of integrals of the higher-dimensional gauge field over $p$-cycles,
\begin{equation}
\theta^i = \int_{\Sigma^{(p)}_i} \Cp.
\end{equation}
However, this is slightly imprecise, because it is only independent of the choice of $\Sigma^{(p)}_i$ in its homology class when $\Cp$ is constructed only from closed forms on $Y$.

\subsection{Masses for extra-dimensional axions}

So far, we have seen that the theory of an extra-dimensional $p$-form gauge field $\Cp$ gives rise to 4d axions that are massless, if we only consider the kinetic term. In general, there should also be effects that give rise to a small 4d mass. These must come from terms in the full theory that involve $\Cp$ {\em without} derivatives acting on it. Such terms are highly restricted, by gauge invariance. They come in two types: couplings to objects that carry a $\Cp$ gauge charge, and Chern-Simons interactions of $\Cp$ with itself or with other gauge fields.

\subsubsection{Brane instantons}
\label{subsec:braneinstantons}

Let's begin by considering objects electrically charged under $C^{(p)}$. These are conventionally referred to as $(p-1)$-branes, and have a $p$-dimensional worldvolume. A brane of charge $q$ under $\Cp$ couples via a term $\int_{\Gamma^{(p)}} q\,\Cp$ on the worldvolume ${\Gamma^{(p)}}$. (This is the dynamical version of a Wilson line operator, and is invariant under large ``winding'' gauge transformations for the same reason the Wilson loop is.) To isolate the axion modes in $\Cp$, we use the ansatz~\eqref{eq:Cpansatz}, so this integral is only nonzero if $\Gamma^{(p)}$ extends {\em only} along internal directions within $Y$. But this means that it is completely localized in 4d spacetime $X$, i.e., it is an {\em instanton} from the viewpoint of the 4d theory. These are often referred to as ``Euclidean brane instantons,'' because the $(p-1)$-brane worldvolume has been wrapped on a cycle $\Gamma^{(p)}$ in $Y$, and only has spatial (Euclidean) directions and does not extend in the time direction. 

How large are the effects of a Euclidean brane instanton? For this we need to consider not just the coupling to $\Cp$ but the other leading term in the brane worldvolume action, namely, the {\em tension} of the brane, which gives a contribution to the action that is linear in the volume:
\begin{equation}
S_\textrm{brane} = \int_{\Gamma^{(p)}} \left( \cT_{(p)} \,{\star_{\Gamma}}1 + \iu q\,\Cp\right). 
\end{equation}
Here $\star_{\Gamma} 1$ is the volume form on the submanifold $\Gamma^{(p)}$, and $\cT_{(p)}$ is the brane's tension.  The $\iu$ comes from a Euclidean continuation. A dilute gas of these Euclidean brane instantons then gives rise to an effective axion potential in the infrared. Representing the cycle in terms of our basis of $p$-cycles as $[\Gamma^{(p)}] = \sum_j m_j [\Sigma^{(p)}_j]$, the simplest estimate is
\begin{equation} \label{eq:Veffbrane}
V_\textrm{eff}^{\textrm{(brane)}}(\{\theta^i\}) =  \left(M^4 \E^{-\cT_{(p)} \,\mathrm{Vol}(\Gamma^{(p)})} \E^{\iu q m_j \theta^j} + \mathrm{h.c.}\right),
\end{equation}
where $M$ has units of mass, and one's first expectation might be that it corresponds to an ultraviolet cutoff energy like the compactification scale at which the $d$-dimensional theory matches to a 4d effective theory of axions. In fact, there could be further suppression: for instance, in a supersymmetric theory the Euclidean brane instantons may generate an effective superpotential or an effective K\"ahler potential, and the eventual size of the potential could then be suppressed by supersymmetry-breaking parameters. Even without supersymmetry, there may be fermion zero modes associated with these instantons, so that what is generated is an 't~Hooft vertex rather than a potential. Independent of such subtleties, we expect to always have the exponential suppression by the brane tension times the volume of the cycle the brane wraps, which we refer to as the instanton action:
\begin{equation} \label{eq:Sinst}
S_\mathrm{inst} = \cT_{(p)} \, \mathrm{Vol}(\Gamma^{(p)}).
\end{equation}
The exponential suppression of the axion potential is very powerful. It implies that, if the volumes are somewhat large in units of whatever physical scale sets the brane tension, these contributions to the axion potential can be very small. Any effect that can generate an axion potential must see the entire cycle over which $\Cp$ is integrated to obtain a given axion mode, and such nonlocal effects are naturally suppressed. In this way, extra-dimensional axion models effectively take the logarithm of the axion quality problem, compared to models of axions as 4d pseudo-Nambu-Goldstone bosons where one finds power-law suppression factors instead of exponential suppression factors.

\subsubsection{Gauge theory instantons}
\label{subsec:gaugeinstantons}

The axion that we're most interested in is the QCD axion, with coupling~\eqref{eq:axiongluon} to gluons. More generally, axions might have similar couplings to any 4d gauge fields. Unlike in models of axions as 4d pseudo-Nambu-Goldstone bosons, where such couplings generally arise from integrating out fermions at one loop, in extra-dimensional models we expect them to arise from dimensional reduction of higher-dimensional Chern-Simons couplings.

Specifically, we consider that the 4d gauge fields are zero modes of higher-dimensional gauge fields, so that (for example) the 4d gluons arise from an SU(3) gauge theory in $4+p$ dimensions. We do not necessarily assume that $d = 4+p$; rather, these gauge fields can be localized on $(p+3)$-branes wrapping a $p$-cycle $\Gamma^{(p)}$ within $Y$, with $[\Gamma^{(p)}] = \sum_j m_j [\Sigma^{(p)}_j]$. Along these branes, we have an action defined on the manifold $N = X \times_w \Gamma^{(p)}$
\begin{equation} \label{eq:braneaction}
S_{p+3} = \int_{N} \left(-\frac{1}{2 g_p^2} \tr[G(x,y) \wedge \star_N G(x,y)] + \frac{k_G}{8\cpi^2} \Cp(x,y) \wedge \tr[G(x,y) \wedge G(x,y)]\right).
\end{equation}
Here $G(x,y) = \dif A(x,y)$ is the 2-form field strength of an SU(3) gauge field and $g_p$ is a gauge coupling of mass dimension $-p/2$. The Chern-Simons term is not gauge invariant; the requirement that $\exp(\iu S_{p+3})$ is well-defined under $\Cp$ gauge transformations implies that $k_G \in \ZZ$. 

The 4d gluon fields arise by simply taking $G(x,y) = G(x)$ to be a $y$-independent, 4d two-form field strength. From~\eqref{eq:hodgeproduct} with $q = 2$, we see that warping plays no role in this case. We obtain the 4d action
\begin{equation}
S_\text{gauge} = \int_X \left(-\frac{1}{2g^2} \tr[G(x) \wedge \star_X G(x)] + \frac{k_G}{8\cpi^2} \sum_j m_j \theta^j(x) \tr[G(x) \wedge G(x)]\right),
\end{equation}
where
\begin{equation} \label{eq:4dgaugecoup}
\frac{1}{g^2} = \frac{1}{g_p^2} \int_{\Gamma^{(p)}} \star_N 1 = \frac{1}{g_p^2} \mathrm{Vol}(\Gamma).
\end{equation}
Thus, we see that the higher-dimensional Chern-Simons coupling gives rise to (properly quantized) couplings of axions to gauge fields in 4d. As usual, these gauge fields then contribute to an axion potential via exponentially small 4d instanton effects, which take the form
\begin{equation} \label{eq:Veffgauge}
V_\textrm{eff}^{\textrm{(gauge)}}(\{\theta^i\}) =  \left(M^4 \E^{-8\cpi^2/g^2} \E^{\iu k m_j \theta^j} + \mathrm{h.c.}\right).
\end{equation}
This is very similar to~\eqref{eq:Veffbrane}, with the brane charge $q$ replaced by the Chern-Simons level $k_G$. The similarity is even stronger if we take into account the volume scaling of $1/g^2$ in~\eqref{eq:4dgaugecoup}, so that we can identify the brane tension $\cT$ with $8\cpi^2/g_p^2$. 

This is no accident. In fact, BPST Yang-Mills instanton solutions lift to $(4+p)$ dimensions, where they are not localized in spacetime but are extended, dynamical objects with $p$-dimensional worldvolumes (taking the solution to be constant along $p$ directions), and these indeed have tension $\cT = 8\cpi^2/g_p^2$. To obtain the 4d instanton, we wrap the worldvolume of such an object over the cycle $\Gamma^{(p)}$. In this sense, the Yang-Mills instantons and the Euclidean brane instantons discussed in \S\ref{subsec:braneinstantons} are not so distinct. Indeed, in string-theoretic examples it is known that small Yang-Mills instantons are precisely the same objects as Euclidean branes~\cite{Witten:1995gx, Douglas:1995bn, Douglas:1996uz}.
 

Of course, once we match to the 4d theory, we have the usual infrared physics of QCD. At low energies the description in terms of a dilute gas of instantons breaks down as the instantons become large and overlapping, so~\eqref{eq:Veffgauge} is no longer accurate. In this regime, one can compute the effective axion potential, which has $O(1)$ deviations in shape from a cosine potential, using the chiral Lagrangian~\cite{DiVecchia:1980yfw}.

\subsubsection{Tree-level masses}
\label{subsec:treelevelmass}

So far everything has worked nicely: higher-dimensional gauge fields automatically give rise to 4d axions. These can have the necessary couplings to 4d gauge fields to solve the Strong CP problem (or to play a role in other phenomenologically interesting scenarios), arising from higher-dimensional Chern-Simons terms. Euclidean branes can give rise to additional contributions to axion potentials, but these are exponentially small and in some cases are in fact identical to small gauge theory instantons, so the axion quality problem is inherently mild.

However, if an axion has a different type of mass term that is {\em not} exponentially small, this would give the axion a large mass, decoupling it from the low-energy theory and eliminating the solution to the Strong CP problem. Because the axion originates from a higher-dimensional gauge field, it is relatively protected against such effects, but there are two mechanisms by which it could obtain a tree-level mass. We summarize these by saying the axion can eat or be eaten.

The first case is a monodromy mass for the axion~\cite{Silverstein:2008sg, McAllister:2008hb, Kaloper:2008fb, Kaloper:2011jz} (also see~\cite{Kallosh:1995hi, Gabadadze:1999na, Gabadadze:2000vw, Gabadadze:2002ff, Dvali:2005an} for a discussion of the standard axion potential in a similar formalism). In four dimensions, it comes from a coupling of the form $\frac{n}{2\cpi} \theta F^{(4)}$, where $F^{(4)} = \dif C^{(3)}$ is the four-form field strength of a 3-form gauge field (which has no propagating degrees of freedom). If $F^{(4)}$ has a standard quadratic kinetic term, this is equivalent to a quadratic mass term $\frac{1}{2} m^2 f^2 \theta^2$. Such a term appears to violate the $\theta \mapsto \theta + 2\cpi$ gauge invariance. In fact, what happens is that the potential has infinitely many branches, related by monodromy. One can dualize $F^{(4)}$ to a ``zero-form field strength'' which is simply an integer $n$; the potential is actually a function of the combination $(\theta - 2\cpi n)$, with a gauge invariance $\theta \mapsto \theta + 2\cpi, n \mapsto n +1$. This is essentially just the familiar physics of a particle on a circle in quantum mechanics. We could say that the axion eats the integer (or the fictitious $(-1)$-form field whose field strength is the integer) to become massive. Alternatively, in the magnetic frame, the $B^{(2)}$ field dual to the axion is eaten by $C^{(3)}$.

The monodromy mass is the axion analogue of a $BF$ term. It could arise from a term of the form $\frac{n}{2\cpi} \Cp \wedge \dif B^{(3-p+m)}$ integrated over $m \geq p$ extra dimensions. This is just another type of Chern-Simons term, but involving two gauge field factors instead of one gauge field factor integrated over a charged object's worldline (in \S\ref{subsec:braneinstantons}) or three gauge field factors (in \S\ref{subsec:gaugeinstantons}). Such a term could also arise from a cubic Chern-Simons term of the form $\frac{1}{4\cpi^2} \Cp \wedge \dif B^{(3-p+m)} \wedge \dif A^{(q)}$ when $A^{(q)}$ has flux over $q+1$ additional extra dimensions, or similarly from Chern-Simons terms with even more fields and more fluxes.

The second case is a Stueckelberg mass for an ordinary 1-form gauge field, which eats an axion to become massive. This mass originates from the axion kinetic term proportional to $|\rmd \theta - q A^{(1)}|^2$, where $q \in \ZZ$ is a $\Uone$ charge associated with the gauge transformation $A^{(1)} \mapsto A^{(1)} + \rmd \lambda$, $\theta \mapsto \theta + q \lambda$. This is dual to a $BF$ term of the form $\frac{q}{2\cpi} B^{(2)} \wedge \rmd A^{(1)}$, with $B^{(2)}$ the 2-form gauge field dual to the axion. This could again be dualized to a frame in which the field $B^{(2)}$ eats the magnetic dual gauge field $\widetilde{A}^{(1)}$. 

Because we are interested in light axions, we will assume that the fields we focus on do not obtain tree-level masses of either type. This is natural, since such a mass is proportional to an integer that can simply be zero (and is not renormalized). In \S\ref{subsec:treelevelagain}, we will revisit the possibility of tree-level axion masses and discuss their relationship with chiral fermions, making more model-dependent assumptions.

\subsection{Axion strings for extra-dimensional axions}
\label{subsec:axionstrings}

Up to this point, the higher-dimensional physics we have discussed gives rise to an effective field theory of light axions just like that in models of axions as 4d pseudo-Nambu-Goldstone bosons. However, the physics at higher energies can be dramatically different in these different types of models. One key difference lies in the physics of axion strings, which are dynamical objects such that the axion field $\theta(x)$ winds around its target space as the spatial position $x$ winds around the string. In a weakly-coupled 4d model like the classic KSVZ or DFSZ models, an axion string can be constructed as a classical solution to the equations of motion, with the classical Peccei-Quinn symmetry restored in the core of the string. For extra-dimensional axions, there is no 4d classical Peccei-Quinn symmetry, and correspondingly the axion string's core probes deep UV physics, as recently emphasized in~\cite{Dolan:2017vmn, Reece:2018zvv, Lanza:2020qmt, Lanza:2021udy, March-Russell:2021zfq, Cicoli:2022fzy, Benabou:2023npn}.

An axion string carries a magnetic charge under the axion, in the sense that $\oint \dif \theta = 2\cpi n$ where the integral is over a circle that links with the string worldsheet, just as $\int_\Sigma F = 2\cpi n$ for an ordinary gauge field strength integrated over a sphere linking a magnetic monopole worldline. In fact, an axion string originates as a magnetic monopole for $\Cp$. Specifically, a magnetic monopole for a $p$-form gauge field in $d$ spacetime dimensions is an object with a $(d - p  - 2)$-dimensional worldvolume. In the compactified theory, we can wrap this worldvolume over $n - p$ internal dimensions to obtain a string in 4d. 

In the extra-dimensional theory, then, the magnetic $(d-p-3)$ brane wraps an $(n-p)$-dimensional submanifold $\Gamma^{(n-p)}$ and extends along a worldsheet $\Lambda \subset X$. Poincar\'e duality associates the homology class $[\Gamma^{(n-p)}] \in H_{n-p}(Y, \ZZ)$ with a cohomology class $[\gamma^{(p)}] \in H^p(Y, \ZZ)$. This is defined so that, for any closed $(n-p)$-form $\xi^{(n-p)}$, we have $\int_Y \gamma^{(p)} \wedge \xi^{(n-p)} = \int_{\Gamma^{(n-p)}} \xi^{(n-p)}$. (Loosely speaking, it can be thought of as a Dirac delta function localizing fields to $\Gamma^{(n-p)}$.) If we decompose the Poincar\'e dual class in our chosen basis, $[\gamma^{(p)}]  = \sum_i n^i [\omega^{(p)}_i]$, then in 4d we have an axion string with
\begin{equation}
\oint_C \dif \theta^i = 2\cpi n^i \,\mathrm{Link}(C, \Lambda),
\end{equation}
where $\mathrm{Link}(C, \Lambda) \in \ZZ$ is the linking number of a closed curve $C$ with the worldsheet $\Lambda$. This means that $\theta^i$ winds $n^i$ times around the string. In particular, to obtain a minimally-charged axion string for which a particular axion $\theta^i$ winds once around the string, we should wrap the magnetic brane on the cycle Poincar\'e dual to $[\omega^{(p)}_i]$. By~\eqref{eq:pformbasis}, this is a cycle that has intersection number $\delta^i_j$ with the cycle $[\Sigma^{(n-p)}_j]$.


An axion string has a tension that is determined by the product of the magnetic brane tension in the $d$-dimensional theory and the warped volume of the wrapped cycle:
\begin{equation} \label{eq:axionstringtension}
\cT_{(2)} = \cT_{(d-p-2)} \int_{\Gamma^{(n-p)}} w(y)\, {\star_{\Gamma^{(n-p)}} 1} = \cT_{(d-p-2)} \mathrm{Vol}_w(\Gamma^{(n-p)}),
\end{equation}
where we have defined the ``warped volume'' $\mathrm{Vol}_w$ in the last step. The tension~\eqref{eq:axionstringtension} should be thought of as the tension of the string core. The winding of the axion around the string contributes, at larger distances, an infrared divergent contribution $\cT_\textsc{IR} \approx \cpi f^2 \log(r_\textsc{IR}/r_\mathrm{core})$.

\subsection{Qualitative features of extra-dimensional axions}

Extra-dimensional axions are appealing because they render the axion quality problem much more mild. The QCD contribution to the axion potential is always exponentially small, since the QCD scale arises from dimensional transmutation. In 4d axion models, this competes with effects from PQ-breaking local operators that are suppressed by powers of $f/\Lambda_\textsc{QG}$. Here $\Lambda_\textsc{QG}$ denotes the fundamental cutoff scale at which gravity becomes strong and all global symmetries are badly broken, which is at most the Planck scale. Such models require new gauge symmetries to forbid dangerous operators up to a high dimension (see, e.g.,~\cite{Georgi:1981pu, Lazarides:1985bj, Casas:1987bw, Kamionkowski:1992mf, Holman:1992us, Barr:1992qq, Ghigna:1992iv, Randall:1992ut, Dine:1992vx, Kallosh:1995hi}). In the extra-dimensional case, on the other hand, terms in the axion potential inherently involve nonlocal effects from extended objects stretched across the extra dimensions. These are always exponentially small. This is not to say that the problem has been trivialized. One still must check that the resulting small potential is small {\em enough} that the Strong CP problem is solved; see, e.g., discussions in~\cite{Svrcek:2006yi, Demirtas:2021gsq}.

Because extra-dimensional axion models have no 4d PQ symmetry breaking, the axion in such models is not a PNGB in the traditional sense. There is a symmetry current that creates an axion particle from the vacuum and that generates a shift of the axion field in the low energy theory, but it is simply the operator $\partial_\mu \theta$, which does not arise from a more fundamental current for a symmetry acting linearly on fields. In the higher-dimensional setting, it arises from the field strength of the gauge field that gave rise to the axion. Thus, extra-dimensional axions are only PNGBs in the sense that the photon is a Nambu-Goldstone boson~\cite{Gaiotto:2014kfa}. The symmetry {\em always} acts nonlinearly, by shifting the field.

Because there is no 4d spontaneous PQ symmetry breaking, an extra-dimensional axion is not associated with a 4d cosmological phase transition~\cite{Cicoli:2022fzy, Reece:2023czb, Benabou:2023npn}. The frequently studied post-inflation axion scenario, in which the axion field is randomized at the PQ phase transition and cosmic strings are produced by the Kibble-Zurek mechanism, which later annihilate axion domain walls at the QCD phase transition, simply does not apply to extra-dimensional axions. Instead, these models behave like the conventional ``pre-inflation scenario.'' This is often described as a scenario where the PQ phase transition happened before inflation, but we do not know if there was a pre-inflation epoch in cosmology, nor do we need one. The important point is that at any time that the universe can be approximated as four-dimensional, there is a light axion field in the spectrum of the theory, which is not the phase of a complex scalar that can be trapped at the origin of field space. In cases where it can be paired with a radial saxion or modulus to form a complex scalar field, the point in field space at which the axion decay constant vanishes lies at infinite distance~\cite{Dolan:2017vmn, Reece:2018zvv}, so cosmological dynamics never traps the modulus at such a point. The would-be 4d PQ symmetry is always already broken.\footnote{This does not preclude the possibility that the early universe went through a phase in which cosmological dynamics was intrinsically higher-dimensional, or that cosmic strings could have been produced by a mechanism other than the standard Kibble-Zurek mechanism. Our comments only apply to the conventional 4d post-inflation axion cosmology. I thank Jim Cline for raising this point.}

The existence of axions arising from higher dimensional gauge fields is, as we have seen in \S\ref{subsec:whatisxdaxion}, a topological property of the theory. In particular, there is no need for additional structure like supersymmetry to explain why an extra-dimensional axion is light: it is inherently exponentially light, provided it does not obtain a tree-level mass from the (again, topological) terms discussed in \S\ref{subsec:treelevelmass}. Topological quantities are reassuringly robust, but most of the information that is actually useful for phenomenology is {\em geometric}, not topological: to compute the values of gauge couplings~\eqref{eq:4dgaugecoup}, axion decay constants or kinetic mixing parameters~\eqref{eq:axionkinetic}, instanton actions~\eqref{eq:Sinst}, or axion string tensions~\eqref{eq:axionstringtension}, we need to understand the geometry of the extra dimensions. Geometry is, unfortunately, much more difficult than topology. However, in special cases, the volumes of certain cycles in the higher dimensions are {\em calibrated}: a minimal volume submanifold in a given topological class $H_p(Y, \ZZ)$ has its volume given by integrating a special $p$-form called a calibration over the cycle. This occurs, for example, when $Y$ is a K\"ahler manifold, $p = 2k$ is even, and the $p$-cycle is holomorphic; in that case, the calibration is $J^{\wedge k}$, where $J$ is the K\"ahler form. We will make use of this to study concrete examples in \S\ref{subsec:Cfouraxions} below. Another example is when $Y$ is a Calabi-Yau $n$-fold, $p = n$ is half the real dimension of $Y$, and the $p$-cycle is special Lagrangian; in that case, the calibration is $\mathrm{Re}\,\Omega$, where $\Omega$ is the holomorphic $n$-form on $Y$. Although calibrated cycles are a mathematical tool to make the computation of volumes easier, it turns out that the conditions under which they are known to exist are commonly {\em also} conditions under which compactifications on $Y$ are supersymmetric. This is a motivation for studying supersymmetric theories that is distinct from phenomenological considerations like the hierarchy problem, or field-theoretic arguments for using supersymmetry to control QFTs in a strongly-coupled regime. Here, supersymmetry provides a geometrical toolkit to compute the axion effective action in otherwise intractable extra-dimensional settings.

\section{Estimating a UV cutoff}
\label{sec:uvcutoffestimate}

\begin{figure}[th]
  \centering
  \includegraphics[width=0.65\textwidth]{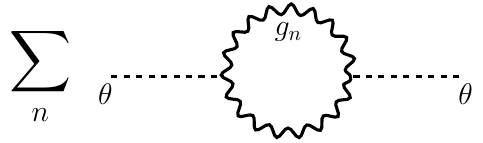}
  \caption{Loop corrections to the axion propagator from the sum over loops of gluon Kaluza-Klein modes.}
  \label{fig:axiongluonloop}
\end{figure}

In this section, we give a simple estimate of an upper bound on the UV cutoff of the theory, above which the extra-dimensional Chern-Simons description breaks down. This claim was advertised, but not explained, at the end of my TASI lecture notes~\cite{Reece:2023czb}; this paper (belatedly) provides the argument. The idea is simply that the axion-gluon coupling is non-renormalizable, and so generally signals a UV cutoff on the theory. We could try to estimate this directly from the higher-dimensional Chern-Simons term appearing in the brane action~\eqref{eq:braneaction}, but this is complicated by the fact that $C^{(p)}$ in general propagates in additional directions transverse to the brane, so the proper way to normalize the field on the brane to read off the cutoff scale is not obvious. Instead, we will work directly with the 4d effective action, but take into account the fact that the axion $\theta$ couples to an entire tower of Kaluza-Klein modes $g_n$ of the gluon. The number of these modes scales like the volume of the $p$-cycle we integrate over to obtain a 4d axion, in units of the cutoff. These modes all contribute a loop correction to the axion propagator, as shown in Fig.~\ref{fig:axiongluonloop}.

We estimate that loops correct the momentum-space propagator of the canonically normalized axion $\hat{\theta}$ as
\begin{equation}
\langle \hat{\theta}(q) \hat{\theta}(-q)\rangle = \frac{1}{q^2 + \iu \varepsilon} \frac{1}{1 + \Pi(q)}, \quad \text{where} \quad \Pi(q) \sim \frac{1}{16\cpi^2} \left(\frac{g^2}{8\cpi^2 f}\right)^2 \,N_\textsc{KK}(q)\, q^2 + \text{other}.
\end{equation}
Here $N_\textsc{KK}(q)$ denotes the number of Kaluza-Klein modes with mass below the scale $q$, and ``other'' refers to any other effects that we have not included by summing over gluon loops. At the UV cutoff $\Lambda$, we should have $\Pi(\Lambda) \sim 1$, so in particular (barring fine-tuned cancellations), we expect that the piece we have estimated should be $\lesssim 1$. From matching the higher-dimensional theory to the 4d theory, we expect that $N_\textsc{KK}(q) \sim 1/g^2$, and hence we obtain a bound of the form $\Lambda \lesssim f/g$. Let us work this out more carefully.

We assume that the gauge kinetic term in $p+4$ dimensions scales like an appropriate power of the cutoff, times a prefactor $\zeta$ that may be large but which we might expect to be typically of order one:
\begin{equation}
\frac{1}{g_p^2} = \zeta \Lambda^p.
\end{equation}
For example, for a gauge coupling on a D-brane in string theory, if we identify $\Lambda$ with the string scale $M_s$, then $\zeta = \frac{1}{(2\cpi)^{p+1} g_s}$. According to~\eqref{eq:4dgaugecoup}, the 4d gauge coupling is given by $1/g^2 = \mathrm{Vol}(\Gamma)/g_p^2$. The number of KK  modes on a $p$-cycle up to a scale $q$ is given by Weyl's law, $N_\textsc{KK}(q) \sim \frac{1}{(4\cpi)^{p/2} \Gamma(\frac{p}{2}+1)} \mathrm{Vol}(\Gamma) q^p$. Putting the pieces together, we obtain a cutoff
\begin{equation} \label{eq:cutoffupperbound}
\Lambda \lesssim \frac{f}{2g} \left[\zeta (4 \cpi)^{6 + p/2} \Gamma(1+p/2)\right]^{1/2}.
\end{equation}
Although this was a fairly crude estimate, the point that we would like to emphasize is that it is proportional to $f/g$, with a proportionality constant that we might expect to be $O(1)$ in many models. For example, in the string theory setting, the proportionality factor $\zeta$ could be made very large only at the expense of going to very small $g_s$ (which makes string modes much lighter than the Planck scale). More generally, small $g_p$ is expected to correlate with a low cutoff on the higher-dimensional theory, due to the Weak Gravity Conjecture~\cite{Arkani-Hamed:2006emk, Harlow:2022gzl}. 

This UV cutoff is one at which the Chern-Simons description of the axion-gluon interaction breaks down. In particular, it is not simply associated with a breakdown of the 4d EFT and the need for a higher-dimensional description. We expect that it is a fundamental ultraviolet scale of the theory. We will see below that in many string theory examples, a cutoff arises at or below a scale $O(1) f/g$, {\em independent} of $g_s$, so the argument we have given here is likely to be overly conservative. In particular, we will see that the scale $O(1) f/g$ has a  concrete physical interpretation: it is the mass scale at which excitations of the axion string appear.

\section{Axion string tension scaling with decay constant}
\label{sec:arguments}

In this section, we will argue that it is reasonable to expect the scale corresponding to the axion string tension to be near, but slightly above, that of the decay constant. More precisely, in a variety of models there is an approximate relation of the form
\begin{equation} \label{eq:TSfsq}
\cT_{(2)} \approx 2\cpi\, S_\mathrm{inst}\, f^2.
\end{equation}
Here we focus on a single axion; $\cT_{(2)}$ is the tension of a minimal axion string, $S_\mathrm{inst}$ is the action of a single instanton for $\theta$ (equal to $8\cpi^2/g^2$ when it arises as a Yang-Mills instanton for a gauge theory with coupling $g$), and $f$ is the axion decay constant. 

A few words about the interpretation of the estimate~\eqref{eq:TSfsq} are in order. First, the tension of an axion string is infrared divergent, as discussed in \S\ref{subsec:axionstrings}. The estimate refers only to the tension of the core of the string. In most of the examples we will study, this arises as the (well-defined) tension of a higher-dimensional brane, multiplied by the volume of a cycle wrapped by that brane. Second, the $\approx$ sign here is somewhat ambiguous. In a very wide range of examples, it is really a $\sim$, that is, the estimate~\eqref{eq:TSfsq} holds up to an $O(1)$ factor. However, in many examples, it is even better: it holds exactly, at least at leading order in various weak-coupling or large-volume limits, due to BPS conditions in the higher-dimensional theory giving rise to the 4d axion theory.

This estimate generally lies in between two other guesses one might have. The first, based on the intuition from 4d field theory axions, is that $\cT_{(2)} \sim \cpi f^2$. This sets the coefficient of the infrared logarithm, so it would make little sense for $\cT_{(2)}$ to be taken to be smaller than this. The second is that $\cT_{(2)} \sim 2\cpi f M_\mathrm{Pl}$. The latter estimate is motivated by the magnetic axion Weak Gravity Conjecture, which holds that $\cT_{(2)} \lesssim 2\cpi f M_\mathrm{Pl}$~\cite{Arkani-Hamed:2006emk}.\footnote{The axion WGC is somewhat degenerate, from the viewpoint of the black hole extremality arguments originally motivating the WGC. However, an argument based on black hole physics for the necessity of cosmic strings in quantum gravity with an axion was given in~\cite{Hebecker:2017uix, Montero:2017yja}.} This bound is in fact saturated in many simple examples (we will discuss some in \S\ref{sec:examples}), which has led to suggestions that the estimate $\cT_{(2)} \sim 2\cpi f M_\mathrm{Pl}$ works quite generally for extra-dimensional axions, at least in the absence of warping (see, e.g., the recent~\cite{Benabou:2023npn}). However, we will also see examples where it fails, whereas~\eqref{eq:TSfsq} holds up in a wider range of examples.

Clearly, we would expect any general relationship among axion string tensions, decay constants, and instanton actions to be modified in a setting with multiple axions that mix with each other through an off-diagonal kinetic matrix $\kappa_{ij}$ as in~\eqref{eq:axionkinetic}. In this section, we focus on a single axion, hoping that the QCD axion is relatively isolated from mixing effects in theories that solve the Strong CP problem. In~\S\ref{subsec:Cfouraxions} below, we will encounter a class of multi-axion examples in which an axion string around which $\theta^i$ winds $w^i$ times has tension $2\cpi w^i \kappa_{ij} S^j$. This is one candidate for a more general estimate in the multi-axion case. Interestingly, in the examples that we study in~\S\ref{sec:examples}, if we read off the value of $f$ and $S_\mathrm{inst}$ from any single axion, we find an axion string with tension of order~\eqref{eq:TSfsq}, although it may not have minimal winding under the axion of interest and may also have nonzero winding of additional axions. This is reminiscent of known 4d field theory examples, where the minimal-tension axion string is often not the minimal-winding axion string (see, e.g.,~\cite{Barr:1992qq, Lu:2023ayc}).

We will now discuss several general, but heuristic, arguments for why we might expect~\eqref{eq:TSfsq} to hold. In \S\ref{sec:examples}, we consider specific UV complete examples (in string theory) where we can explicitly verify the statement.

\subsection{Scaling of decay constants with volumes}
\label{subsec:scalingwithvolume}

Let's first consider a simple geometrical argument. Suppose that physics in our $d$-dimensional theory is dominated by a single length scale $\ell$ (e.g., the string scale, in string theory, or the 11d Planck scale in M-theory). We will also suppose that warping is not important, that dimensionless moduli like $\Phi$ in extra dimensions take order-one values, and that all higher-dimensional couplings and tensions like $e_p$ or $\cT_{(p)}$ are order-one in units of $\ell$. We are not claiming these assumptions are universally valid, just using them as a starting point to explore our expectations. 

We focus on a single axion $\theta$ arising from $\Cp$ dimensionally reduced on a $p$-cycle $[\Sigma^{(p)}]$, corresponding to a warped harmonic form ${\widehat \omega}^{(p)}$, and intersecting a dual cycle $[\Gamma^{(n-p)}]$. Then our formulas from \S\ref{sec:xdaxions} imply:
\begin{align}
\cT_{(2)} &\sim \ell^{-2+p-n}\, \mathrm{Vol}(\Gamma^{(n-p)}), \nonumber \\
S_\mathrm{inst} &\sim \ell^{-p}\, \mathrm{Vol}(\Sigma^{(p)}), \nonumber \\
f^2 &\sim \ell^{2p - n - 2} \int_Y {\widehat \omega}^{(p)} \wedge \star_Y {\widehat \omega}^{(p)}.
\end{align}
Here $\Sigma^{(p)}$ and $\Gamma^{(n-p)}$ are actual submanifolds, not just homology classes; we expect that these are (at least locally) volume-minimizing representatives of the class. However, at the level of the current scaling argument, such details are not very important. Now, ${\widehat \omega}^{(p)}$ is dimensionless, whereas $\star {\widehat \omega}^{(p)}$ has mass dimension $2p - n$. By Poincar\'e duality, we have
\begin{equation}
\int_Y {\widehat \omega}^{(p)} \wedge \star_Y {\widehat \omega}^{(p)} = \int_{\Gamma^{(n-p)}} \star_Y {\widehat \omega}^{(p)}.
\end{equation}
How can we estimate this? Assuming that no curvatures are particularly large, we can make a crude assumption that the answer is not very sensitive to our precise choice of representative from a cohomology class. Thus, we think of $[{\widehat \omega}^{(p)}]$ as a function localized on ${\Gamma^{(n-p)}}$, smeared along the $p$ transverse directions and multiplied by their differentials $\dif x^1 \wedge \cdots \wedge \dif x^p$, such that the total weight is $1$ when integrated over $\Sigma^{(p)}$. Then $\star_Y {\widehat \omega}^{(p)}$ looks like the same localized, smeared function, but now with the differentials {\em along} $\Gamma^{(n-p)}$, $\dif x^{p+1} \wedge \cdots \wedge \dif x^n$. Thus, its integral now acquires a factor of the volume of $\Gamma^{(n-p)}$, but is suppressed by the volume of $\Sigma^{(p)}$ because we are no longer integrating over the transverse directions ${\widehat \omega}^{(p)}$ was smeared over. Thus, we estimate
\begin{equation} \label{eq:omegasqintegral}
\int_Y {\widehat \omega}^{(p)} \wedge \star_Y {\widehat \omega}^{(p)} \sim \frac{\mathrm{Vol}(\Gamma^{(n-p)})}{\mathrm{Vol}(\Sigma^{(p)})}.
\end{equation}
We will see that, despite the blithe assumptions that went into it, this estimate actually works in at least a significant class of examples.\footnote{In some cases, such as product manifolds, it can be exact; examples are discussed in, e.g.,~\cite{Svrcek:2006yi}. In some cases in the literature an estimate of the form $\mathrm{Vol}(Y)/\mathrm{Vol}(\Sigma^{(p)})^2$ is assumed, implicitly or explicitly; this estimate agrees with~\eqref{eq:omegasqintegral} for product manifolds (or fibrations) but can be much less accurate in other cases.} It implies
\begin{equation}
\cT_{(2)} \sim S_\mathrm{inst}\, f^2.
\end{equation}
This argument is (clearly) too crude to produce a factor like $2\cpi$. This factor will be better motivated by other arguments and examples below. Although we have omitted warping from our argument, it is plausible that it does not change the relationship because it affects the left- and right-hand sides in compatible ways. The warping enters via a single factor of $w(y)$ in the integrand defining $\cT_{(2)}$ in~\eqref{eq:axionstringtension}, and also in the integrand defining $f^2$ in~\eqref{eq:axionkinetic}. In the simplest example of a warped interval, this expectation holds up because the axion wavefunction peaks in the region of greatest warping (smallest $w(y)$) and the axion string falls into this region as well; see~\cite{Benabou:2023npn} for calculations in this case.

\subsection{Bogomolnyi bounds and repulsive forces}

One way to try to quantify the instanton action and axion string tension is to perform an explicit semiclassical calculation within the low-energy effective theory. For an axion minimally coupled to gravity, the corresponding Euclidean instanton solutions~\cite{Giddings:1987cg, Coleman:1989zu} and axion string solutions~\cite{Cohen:1988sg} are well-studied, but singular, and the solutions are not expected to capture the physics of the core of the object accurately. In the core of both objects, a scalar radial mode or {\em saxion} field is expected to be excited in UV completions. In our examples, such moduli generally arise as modes of the metric in the extra dimensions, which determine the volumes of the cycles used to compute the instanton action, as in~\eqref{eq:Sinst}, and the axion string tension, as in~\eqref{eq:axionstringtension}. The coupled equations for the axion-saxion system are often better behaved, and in any case, capture the physics of the instanton or string core more accurately. In this subsection, we will study classical solutions to the equations of motion for an axion together with a saxion $\phi$, coupled through a $\phi$-dependent axion decay constant $f(\phi)$. (See also~\cite{Rudelius:2024vmc} for some related recent work.)

\subsubsection{Euclidean instantons}

The instanton solutions that we study are Euclidean solutions to the coupled equations for the scalar modulus $\phi$ and the two-form gauge field $B^{(2)}$ (with field strength $H^{(3)}$) dual to the axion, with action
\begin{equation}
I_E = \int \left(\frac{1}{2} g(\phi)^2 H^{(3)} \wedge \star H^{(3)} + \frac{1}{2} h(\phi)^2 \rmd \phi \wedge \star \rmd \phi \right).
\end{equation}
We take $g(\phi), h(\phi) > 0$. We denote the asymptotic value of the field $\phi$ at $r \to \infty$ by $\phi_*$. Then we read off from the action that the axion decay constant is 
\begin{equation}
f = \frac{1}{2\cpi g(\phi_*)}.
\end{equation} 
An instanton solution is magnetically charged under $B^{(2)}$, so it has (for minimal instanton number)
\begin{equation}
\int_{S^3} H^{(3)} = 2\cpi,
\end{equation}
when integrated over any 3-sphere surrounding the instanton core. We can find a minimum-action solution using the Bogomolnyi trick:
\begin{equation}
I_E = \frac{1}{2} \int \left| h(\phi) \rmd \phi \mp g(\phi) {\star H^{(3)}} \right|^2 \pm \int h(\phi) g(\phi) \rmd \phi \int_{S^3} H^{(3)}.
\end{equation}
In the last term, $\int_{S^3} H^{(3)}$ is a topological invariant and $\int h(\phi) g(\phi) \rmd \phi$ depends only on the boundary values $\phi_0 = \lim_{r \to 0} \phi(r)$ and $\phi_* = \lim_{r \to \infty} \phi(r)$. Provided that solutions have the property that they always tend to a fixed value of $\phi_0$ (typically $0$, $+\infty$, or $-\infty$), we can search for a solution by minimizing the first term.
Thus, we obtain an instanton solution by solving
\begin{equation}
h(\phi)\rmd \phi = \pm g(\phi) {\star H^{(3)}} = \mp \frac{g(\phi)}{\cpi r^3} \rmd r,
\end{equation}
with the sign chosen to obtain a well-behaved solution (depending on the choice of $g$ and $h$). 
We introduce the quantity 
\begin{equation}
S(\phi) = 2\cpi \int^{\phi}_{\phi_0} h(\phi') g(\phi') \rmd \phi',
\end{equation}
and then we see that for the Bogomolnyi solution we have
\begin{equation} \label{eq:Sphi}
S_\mathrm{inst} = |S(\phi_*)| = 2\cpi \left|\int^{\phi_*}_{\phi_0} h(\phi) g(\phi) \rmd \phi\right|.
\end{equation}

Remarkably, these solutions persist when we include gravity. Without $\phi$, there are gravitational instanton solutions (solutions to the coupled equations for the metric and the axion) that are singular in their core. With a scalar modulus $\phi$, gravitational solutions can be instantons or two-sided wormholes, and there is a special case---the {\em extremal} gravitational instanton---for which the metric is exactly flat. To find such solutions, we should either use the action involving the field $B^{(2)}$, or we must use a wrong-sign axion kinetic term (or allow imaginary axion field values). There is a large literature on this topic; some key references include~\cite{Giddings:1987cg, Coleman:1989zu, Bergshoeff:2004fq, Bergshoeff:2004pg, Gutperle:2002km, Arkani-Hamed:2007cpn} and recent discussions in the context of the Weak Gravity Conjecture include~\cite{Heidenreich:2015nta, Hebecker:2018ofv}. While there is some debate as to the correct order-one coefficient in the axion WGC, one plausible answer is that it is determined by the extremal instantons with flat metric, i.e., by the action $S(\phi_*)$ determined by~\eqref{eq:Sphi}.

\subsubsection{Axion strings}

We can find an axion string solution in the limit when gravity is negligible. We work in coordinates where the vortex worldvolume stretches along directions $t, y$, the axion winds around an angular direction $\varphi$, and the radial distance from the axion core is measured by $r$. The metric, then, is 
\begin{equation} \label{eq:stringmetric}
\rmd s^2 = -\rmd t^2 + \rmd y^2 + \rmd r^2 + r^2 \rmd \varphi^2.
\end{equation}
The solution carries magnetic charge under the axion $\theta$, with $\theta = \varphi$. In this case, we work directly with $\theta$ instead of its dual $B$-field, and the kinetic term depends on the function $f(\phi) \equiv 2\cpi/g(\phi)$. The directions $t, y$ play no role in the solutions we are interested in, so we can factor them out: we introduce the notation $\ast$ for the Hodge star in the $(r, \varphi)$ directions only, and then
\begin{equation}
I_E = -\int \rmd t  \wedge \rmd y \wedge \left[\frac{1}{2} f(\phi)^2 \rmd \theta \wedge {\ast \rmd \theta} + \frac{1}{2} h(\phi)^2 \rmd \phi \wedge {\ast \rmd \phi}\right] \equiv \int \rmd t \wedge \rmd y \wedge {\cal I},
\end{equation}
and we can apply the Bogomolnyi trick to ${\cal I}$:
\begin{equation}
{\cal I} = -\left[\frac{1}{2} \int |h(\phi) \rmd \phi \mp f(\phi) {\ast \rmd \theta}|^2 \pm \int h(\phi) f(\phi) \rmd\phi \int_{S^1} \rmd \theta\right].
\end{equation} 
Again, the trick is based on the fact that $\int_{S^1} \rmd \theta$ is a topological invariant while $\int h(\phi) f(\phi) \rmd\phi$ depends only on the boundary values of $\phi$, so that we can focus on minimizing the first term.
We obtain the axion string solution by solving
\begin{equation}
h(\phi) \rmd \phi = \pm f(\phi) {\ast \rmd \theta} = \mp \frac{f(\phi)}{r} \rmd r,
\end{equation}
choosing the sign so that the solution is well-behaved in the core of the string. Due to the infrared divergence associated with axion strings, the solution will generically {\em not} be well-behaved as $r \to \infty$. Instead, we impose an infrared cutoff at some value $r_*$ with a chosen $\phi(r_*) = \phi_*$, and take the tension computed from the solution with $0 \leq r \leq r_*$ to be the ``core tension'' ${\cal T}_{(2)}$. We define
\begin{equation}
\mathcal{T}(\phi) = 2\cpi \int_{\phi_0}^{\phi} h(\phi')f(\phi')\rmd\phi',
\end{equation} 
and so estimate
\begin{equation}
{\cal T}_{(2)} = |\mathcal{T}(\phi_*)| = 2\cpi \left|\int_{\phi_0}^{\phi_*} h(\phi)f(\phi) \rmd\phi\right| = \left|\int_{\phi_0}^{\phi_*} \frac{h(\phi)}{g(\phi)} \rmd \phi\right|.
\end{equation}
Unlike the case of instantons, axion strings coupled to gravity do not admit a flat metric. Various aspects of axion string solutions with gravity have been discussed in, e.g.,~\cite{Cohen:1988sg, Greene:1989ya, Gregory:1996dd, Dolan:2017vmn, Lanza:2021udy}. As we will mention below, in the supersymmetric case they are somewhat better understood.

\subsubsection{Interpretation}

We have derived a Euclidean instanton action and an axion string tension. They are both given by integrals over $\phi$, and the {\em integrands} are related by
\begin{equation} \label{eq:differentialrelation}
\frac{\rmd{\cal T}(\phi)}{\rmd \phi} = \pm \frac{1}{2\cpi g(\phi)^2} \frac{\rmd S(\phi)}{\rmd \phi}.
\end{equation}
The relative minus sign can arise if the solution is such that, for example, $\phi_* < \phi(r \to 0)$ for the instanton solution but $\phi_* > \phi(r \to 0)$ for the string solution, so the orientation of the $\cT(\phi)$ and $S(\phi)$ integrals differs. On the other hand, the relationship~\eqref{eq:TSfsq} is equivalent to
\begin{equation}
|\cT(\phi_*)| \approx \frac{1}{2\cpi g(\phi_*)^2} |S(\phi_*)|.
\end{equation}
Thus, to the extent that the integral of the relation~\eqref{eq:differentialrelation} is dominated near $\phi = \phi_*$, we see that it lends support to~\eqref{eq:TSfsq}.

The relationship~\eqref{eq:differentialrelation} can be understood as a no-force condition. Indeed, it is expected in quantum gravity that there exists an object with any given gauge charge that is (at least marginally) self-repulsive. In other words, for such an object, the repulsive force mediated by the gauge interaction is stronger than the combined force of gravity and attractive scalar forces acting on the object. This Repulsive Force Conjecture (RFC) is one particular variant of the Weak Gravity Conjecture~\cite{Arkani-Hamed:2006emk, Palti:2017elp, Heidenreich:2019zkl}. The RFC refers specifically to the leading-power, long-range force mediated by massless fields. For codimension 2 objects like axion strings, a remarkable simplification arises: gravity itself does {\em not} mediate such a force. This is in contrast to Newtonian gravity, which would lead us to expect a logarithmic potential separating two massive objects. Instead, in general relativity, codimension 2 objects backreact on spacetime to produce a flat space with a deficit angle~\cite{Gott:1982qg}. Thus, the RFC for axion strings simply requires that the repulsive axion-mediated force at least compensate the attractive saxion-mediated force, a condition that (for a single saxion field) becomes
\begin{equation}
(2\cpi f(\phi))^2 \geq \frac{1}{h(\phi)^2} \left(\frac{\partial \cT}{\partial \phi}\right)^2.
\end{equation}
In the case that this inequality is saturated, it together with~\eqref{eq:Sphi} implies exactly~\eqref{eq:differentialrelation}.

\subsubsection{Example: exponential ansatz}
\label{subsubsec:exponentialex}

Suppose that $h(\phi) = 1$ (i.e., a canonically normalized modulus $\phi$) and
\begin{equation}
g(\phi) = g \E^{-b \phi},
\end{equation}
where without loss of generality we take $\phi_* = 0$ and $b > 0$. The instanton solution is given by
\begin{equation}
\E^{b \phi(r)} = 1 + \frac{b g}{2\cpi r^2}.
\end{equation}
This has the desired property that $\phi(r) \to 0$ at $r \to \infty$, whereas $\phi(r) \to \infty$ at $r \to 0$. In particular, in the core of the instanton, $g(\phi) \to 0$. Thus, we find that
\begin{equation}
S_{\mathrm{inst}} = 2\cpi \int_0^\infty g(\phi) \rmd \phi = \frac{2\cpi g}{b}.
\end{equation}

The vortex solution is given by
\begin{equation}
\E^{b \phi(r)} = \frac{2\cpi g}{b \log(r_0/r)}.
\end{equation}
This gives a well-defined, real value of $\phi$ for $r < r_0$. At larger distances the solution doesn't make sense, but we only care about the answer up to some infrared cutoff $r_* < r_0$ at which $\phi = \phi_* = 0$. In the vortex core, $\phi \to -\infty$. Thus we have
\begin{equation}
{\cal T} = \int^0_{-\infty} \frac{1}{g} \E^{b \phi} \rmd \phi = \frac{1}{bg}.
\end{equation}
We have $f = \frac{1}{2\cpi g}$, so
\begin{equation} \label{eq:TSfsqexponentialex}
2\cpi S_\mathrm{inst} f^2 = 2 \cpi \frac{2\cpi g}{b} \left(\frac{1}{2 \cpi g}\right)^2 = \frac{1}{bg} = \cT.
\end{equation}
Thus, the relationship~\eqref{eq:TSfsq} holds precisely, for solutions saturating the Bogomolnyi bound, in these examples.

\subsubsection{Example: SUSY form, power law ansatz}
\label{subsubsec:powerlawex}

In a supersymmetric theory, $\theta$ is partnered with a real modulus $\phi$ in a complex field $\phi - \frac{\iu}{2\cpi} \theta$. Thus, the kinetic terms for $\phi$ and $\theta$ are related via $f(\phi)^2 = \frac{1}{4\cpi^2} h(\phi)^2$. Let us choose the ansatz
\begin{equation}
f(\phi) = M \phi^\gamma  \quad \Rightarrow \quad g(\phi) = \frac{1}{2\cpi M} \phi^{-\gamma}, \quad h(\phi) = 2\cpi f(\phi) = \frac{1}{g(\phi)}.
\end{equation}
The instanton solution with the asymptotic modulus $\phi_*$ is given by:
\begin{equation}
\phi(r) = \begin{cases} 
      \left(\phi_*^{1+2\gamma} + \frac{1}{8 (1+2\gamma)M^2 \cpi^3 r^2}\right)^{1/(1+2\gamma)}, & \gamma > -\frac{1}{2}, \\
      \phi_* \exp\left[-\frac{1}{8 M^2 \cpi^3 r^2}\right], & \gamma = -\frac{1}{2}, \\
      \left(\phi_*^{1+2\gamma} + \frac{1}{8 |1+2\gamma| M^2 \cpi^3 r^2}\right)^{1/(1+2\gamma)}. & \gamma < -\frac{1}{2}.
    \end{cases}
\end{equation}
We have $\phi(r) \to \infty$ as $r \to 0$ for the case $\gamma > -1/2$, and $\phi(r) \to 0$ as $r \to 0$ for $\gamma \leq -1/2$. In particular, this means that $g(\phi) \to 0$ in the instanton core for $\gamma > 0$ and for $\gamma \leq -1/2$, but $g(\phi) \to \infty$ in the instanton core for $-1/2 < \gamma < 0$. 
Because $h(\phi)g(\phi) = 1$ for this example, for cases where $\phi \to 0$ in the instanton core we have
\begin{equation}
S(\phi_*) = 2\cpi \int^{\phi_*}_0 h(\phi)g(\phi) \rmd \phi = 2\cpi \phi_*.
\end{equation}
This makes sense, since we expect instanton effects to have the form $\exp(-S + \iu \theta)$, and holomorphy dictates that this must match $\exp(-2\cpi(\phi - \iu \theta/(2\cpi)))$. In the cases where $\phi \to \infty$ in the instanton core, we obtain an unbounded action. We conclude that the case $\gamma > -1/2$ is somehow unphysical, and indeed we do not know of UV complete examples where such exponents appear {\em and} the power law ansatz remains valid at large $\phi$; see~\S\ref{subsubsec:swisscheese} for some related discussion. 

 In this case, the axion string solution is simple, as $f(\phi)/h(\phi) = 1/(2\cpi)$ is constant. Hence for the string we have
 \begin{equation}
 \phi(r) = \frac{1}{2\cpi} \log(r_0/r).
 \end{equation}
Notice that this will be true for {\em any} axion string obeying the SUSY ansatz, not just the power law $f(\phi)$. We have $\phi(r) \to \infty$ in the string core. This means that $f(\phi) \to 0$ as $r \to 0$ if $\gamma < 0$, and $f(\phi) \to \infty$ as $r \to 0$ if $\gamma > 0$. The string tension is
\begin{equation}
{\cal T} = \int_{\phi_*}^\infty \left(2\cpi M\right)^2 \phi^{2\gamma} \rmd \phi = 
\frac{(2\cpi M)^2}{|1+2\gamma|} \phi_*^{1+2\gamma},
\end{equation}
if $\gamma < -\frac{1}{2}$. If $\gamma = -1/2$, the integral is logarithmically divergent in the string core. Thus, we focus on the case $\gamma < -\frac{1}{2}$ where both the instanton action and string tension are finite. In this limit, we have
\begin{equation}
2\cpi S_{\mathrm{inst}} f^2 = (2\cpi)^2 \phi_*^{1+2\gamma} M^2 = |1 + 2\gamma| {\cal T}.
\end{equation}
Thus, the relationship~\eqref{eq:TSfsq} holds up to the constant factor $|1 + 2\gamma|$, which is likely to be $O(1)$ in examples. In particular, in the case $\gamma = -1$, the canonically normalized modulus is proportional to $\log \phi$, and in terms of this field we obtain an exponential coupling to the axion as in \S\ref{subsubsec:exponentialex}. In this case $|1 + 2\gamma| = 1$, consistent with our earlier result~\eqref{eq:TSfsqexponentialex}.

In the SUSY context, solutions including dynamical gravity can be found, but they do not have the flat metric~\eqref{eq:stringmetric}. Instead, the plane transverse to the string can be given a complex coordinate $z$ and the metric takes the form $-\rmd t^2  + \rmd y^2 + a(z, \overline{z}) \rmd z\, \rmd \overline{z}$, where $a(z, \overline{z})$ is determined by the K\"ahler potential evaluated on the solution for $\phi$ and $\theta$~\cite{Greene:1989ya}. In this context, the relation~\eqref{eq:differentialrelation} is equivalent to equation (3.7) of~\cite{Lanza:2021udy}. Their analysis singles out the $\gamma = -1$ case (canonically normalized modulus proportional to $\log \phi$) because they focus on scenarios in which the $\phi \to \infty$ limit is physical, and assume that the magnetic axion WGC bound is asymptotically saturated.

\subsection{Magnetic and electric tension scales}
\label{subsec:magneticelectric}
  
We can think of axion strings as magnetically charged objects for axions, and instantons as electrically charged objects. The analogue of ``tension'' for an instanton is the action. Then we can interpret~\eqref{eq:TSfsq} as a relationship between the ratio of magnetic and electric tension scales and the axion coupling constant. It suggests a general statement: given a $p$-form gauge field with coupling constant $g_\mathrm{el}$, whose magnetic dual is a $(d-p-2)$-form gauge field with coupling constant $g_\mathrm{mag} = 2\cpi/g_\mathrm{el}$, the ratio of the tension scale $\cT_\mathrm{mag}$ of magnetically charged objects to the tension scale $\cT_\mathrm{el}$ of electrically charged objects is
\begin{equation} \label{eq:magelratio}
\frac{\cT_\mathrm{mag}}{\cT_\mathrm{el}} \approx \frac{g_\mathrm{mag}}{g_\mathrm{el}}.
\end{equation}
One reason that this statement is plausible is that it is manifestly invariant under exchanging the roles of electric and magnetic quantities. In the axion case, we identify $\cT_\mathrm{mag}$ with the axion string tension, $\cT_\mathrm{el}$ with the instanton action $S$, and $g_\mathrm{el} = 1/f$, and~\eqref{eq:magelratio} is precisely~\eqref{eq:TSfsq}.

This statement is obviously not true in general, so we should take some care to interpret it. We suggest that it might apply in scenarios where both electrically and magnetically charged objects are somehow fundamental, by which mean they are not solitons. It would not apply, for example, to a $\Uone$ gauge theory arising from higgsing of $\SU(2)$, as then the magnetic monopoles are 't~Hooft-Polyakov solitonic solutions whose mass scale depends on the higgsing scale, which can be freely varied. It also does not apply with $\cT_\mathrm{el}$ interpreted as the mass of just any electrically charged object. For example, chiral fermions can carry electric charge and be exactly massless (and in the real world, the electron is nearly massless, relative to fundamental scales); if we identified their mass as the scale $\cal{T}_\mathrm{el}$, equation~\eqref{eq:magelratio} would predict massless magnetically charged objects, which is obviously not the case in general. Instead, we should interpret the scale $\cal{T}_\mathrm{el}$ as the scale at which an infinite tower of electrically charged particles appears. Such towers are expected to always exist, for weakly-coupled gauge theories~\cite{Heidenreich:2015nta, Heidenreich:2016aqi,Montero:2016tif, Andriolo:2018lvp}. In the case of higher-form gauge fields, the electrically charged objects are strings or branes that admit a variety of excitations, and if these objects are not solitons we expect that the tension of a single such object is already a good proxy for the fundamental electric scale. We will also assume that the Yang-Mills instanton action is the correct electric tension scale in the axion case.

A simple motivation for~\eqref{eq:magelratio} comes from considering the classical self-energy of the magnetic monopole. That is, we integrate the energy stored in a monopole's magnetic field down to some UV cutoff radius $r_c$, finding a magnetic monopole tension
\begin{equation}
\cT_\mathrm{mag} \sim \frac{2\cpi}{g_\mathrm{el}^2} \frac{1}{r_c^{p}}.
\end{equation}
If we identify $r_c^{-1}$ with the mass or tension scale associated with a tower of electrically charged objects, this becomes~\eqref{eq:magelratio} with the interpretation of $\cT_\mathrm{el}$ that we have discussed above. Such an identification is reasonable: the tower signals a breakdown of the effective field theory, in a way that a single particle like an electron does not.

The monopole self-energy argument that we have just given bears some resemblance to the Magnetic Weak Gravity Conjecture~\cite{Arkani-Hamed:2006emk, Harlow:2022gzl}. It is worth commenting on how~\eqref{eq:magelratio} relates to the Weak Gravity Conjecture in general. The Weak Gravity Conjecture takes the form:
\begin{equation}  \label{eq:WGC}
\cT_\mathrm{el}^2 \leq \frac{1}{\gamma_\mathrm{el}} g_\mathrm{el}^2 M_\mathrm{Pl}^{D-2}, \quad \cT_\mathrm{mag}^2 \leq \frac{1}{\gamma_\mathrm{mag}} g_\mathrm{mag}^2 M_\mathrm{Pl}^{D-2},
\end{equation}
where the coefficients $\gamma_\mathrm{el}$ and $\gamma_\mathrm{mag}$ are chosen so that large extremal black holes saturate the inequalities, and depend in general on scalar moduli couplings to the gauge-field kinetic term. In most discussions of the WGC, it is assumed that the $\gamma$ factors are order-one numbers, which is the case in the simplest examples. Let us say that a state approximately saturates the naive WGC bound if $\cT^2 \sim g^2 M_\mathrm{Pl}^{D-2}$. Then we see from~\eqref{eq:WGC} that if the electric and magnetic states both approximately saturate their respective naive WGC bound, the ratios of magnetic and electric tensions satisfy~\eqref{eq:magelratio}. This is the case for a large catalogue of standard examples. One way to restate~\eqref{eq:magelratio} is:

\medskip
\noindent {\em If the electric WGC tower appears at an energy scale parametrically below the naive WGC scale, then magnetically charged states also appear parametrically below the naive magnetic WGC scale by the same factor.}\\

One reason that I have phrased this in terms of the {\em naive} WGC scale is that I expect that towers of states {\em always} exist that parametrically saturate the WGC bound~\eqref{eq:WGC}. Extremal black hole states do, by definition, and we expect charged states high in a tower to asymptotically become black holes. Hence, my expectation is that an electrically charged tower appears parametrically below the naive WGC scale precisely in cases where $\gamma_\mathrm{el} \gg 1$, and approximately saturates the WGC once $\gamma_\mathrm{el}$ is taken into account. Thus,~\eqref{eq:magelratio} in the context of the Weak Gravity Conjecture leads me to expect that, in general, $\gamma_\mathrm{el} \sim \gamma_\mathrm{mag}$. It is not obvious how to check this for general scalar couplings to gauge fields, because explicit extremal black hole solutions are not known in the general case. In the special case of exponential couplings of the form $\exp(-\alpha \phi)|F_{p+1}|^2$, we have explicit solutions (dilatonic black holes)~\cite{Gibbons:1982ih, Myers:1986un, Gibbons:1987ps, Garfinkle:1990qj, Horowitz:1991cd}, and it is true that $\gamma_\mathrm{el} = \gamma_\mathrm{mag}$~\cite{Heidenreich:2015nta}. Such exponential couplings are expected to appear in asymptotic regions of the string landscape on quite general grounds~\cite{Ooguri:2006in, Heidenreich:2017sim, Heidenreich:2018kpg, Grimm:2018ohb, Stout:2021ubb, Stout:2022phm}, but with $\gamma$ of $O(1)$. In examples where we expect to find $\gamma \gg 1$, such as for moduli controlling small cycle volumes within a larger compactification, the couplings are power-law rather than exponential and the black hole solutions are not known. Nonetheless, we can check explicit examples of tension scales to see that~\eqref{eq:magelratio} holds. We find such axion examples in~\S\ref{subsubsec:swisscheese} and (with a twist) in~\S\ref{subsubsec:swisscheese1}.

\subsubsection{Magnetic volume scaling argument}

With the argument of this subsection in mind, it is useful to revisit our volume scaling argument from~\S\ref{subsec:scalingwithvolume}. Instead of compactifying $C^{(p)}$ to obtain the axion $\theta$, we could have considered its magnetic dual field $C^{(d-p-2)}_\textsc{M}$. Under dimensional reduction, this gives rise to the 2-form gauge field $B^{(2)}$ dual to $\theta$ by integrating over the dual cycle $[\Gamma^{(n-p)}]$. The prefactor in the kinetic term of $B^{(2)}$ is $1/\left(2\cpi f\right)^2$, and we obtain an estimate:
\begin{equation}
\frac{1}{\left(2\cpi f\right)^2} \propto \int_Y \widehat{\gamma}^{(p)} \wedge \star_Y \widehat{\gamma}^{(p)},
\end{equation}
with $[\gamma^{(p)}]$ as a cohomology class defined in~\S\ref{subsec:axionstrings} and $\widehat{\gamma}^{(p)}$ denoting its warped harmonic representative.
The analogue of~\eqref{eq:omegasqintegral} is then
\begin{equation}
\int_Y \widehat{\gamma}^{(p)} \wedge \star_Y \widehat{\gamma}^{(p)} \sim \frac{\mathrm{Vol}(\Sigma^{(p)})}{\mathrm{Vol}(\Gamma^{(n-p)})},
\end{equation}
which is the reciprocal of the estimate in the electric case.
This shows that the volume scaling estimate~\eqref{eq:omegasqintegral}  is consistent with electric-magnetic duality, whereas other estimates that have appeared in the literature (like $\mathrm{Vol}(Y)/\mathrm{Vol}(\Sigma^{(p)})^2$) are not---they would take on a drastically different form in a dual frame.

\subsubsection{Axion string tension and magnetic monopole scale}

A theory with axions coupled to gauge fields automatically produces a tower of particles electrically charged under the gauge field~\cite{Heidenreich:2021yda}, in the form of closed loops of axion string with excitations of charged modes on the worldsheet (which must exist by anomaly inflow arguments~\cite{Callan:1984sa}). The mass scale $M_\mathrm{el}$ of the tower of electrically charged particles is related to the axion string tension $\cT_{(2)}$ by:
\begin{equation}
M_\mathrm{el} = \sqrt{2\cpi \cT_{(2)}} \approx 2\cpi \sqrt{S_\mathrm{inst}} f,
\end{equation} 
using~\eqref{eq:TSfsq}. If we then take~\eqref{eq:magelratio} seriously, and denote the electric coupling of the gauge theory by $e$, we would expect to find magnetic monopoles at the mass scale
\begin{equation} \label{eq:MmagSinstRel}
M_\mathrm{mag} \approx \frac{2\cpi}{e^2} M_\mathrm{el} \approx \frac{4\cpi^2}{e^2} \sqrt{S_\mathrm{inst}} f.
\end{equation}
In the context of interactions between axions and monopoles, there are two important length scales, the monopole's classical radius $r_c = \cpi/(e^2 M_\mathrm{mag})$ and the axion screening length scale near the monopole core $r_0 = e/(8\cpi^2 f)$~\cite{Fischler:1983sc}. In terms of these scales, we can solve~\eqref{eq:MmagSinstRel} for $S_\mathrm{inst}$ and find the estimate
\begin{equation}
S_\mathrm{inst}^\mathrm{est.} \approx \frac{4\cpi^2}{e^2} \left(\frac{r_0}{r_c}\right)^2.
\end{equation}
This can be compared with the actual computation of the instanton action in the form of a monopole loop with dyonic winding from~\cite{Fan:2021ntg},
\begin{equation}
S_\mathrm{inst} \approx \frac{4\cpi^2}{e^2} \sqrt{\frac{\max(r_c,r_0)}{r_c}}.
\end{equation}
We see that these agree precisely when $r_c = r_0$, i.e., when the axion field is screened at the classical radius of the magnetic monopole. This is a nontrivial consistency requirement on the picture that we have sketched, which need not be true within every effective field theory, but could be a property of UV-complete examples where both the magnetic monopole and the axion string arise from branes wrapped on cycles in extra dimensions. This condition translates to the expectation that
\begin{equation}
M_\mathrm{mag} \approx \frac{8\cpi^3}{e^3} f.
\end{equation}
For example, if $f \sim 10^{12}\,\mathrm{GeV}$ this predicts that the mass of a magnetic monopole is around $10^{16}\,\mathrm{GeV}$.  

\section{Axion string tension in examples}
\label{sec:examples}

In this section, we will show that the expectation~\eqref{eq:TSfsq} holds in several examples in string theory. In each example, we will also evaluate the ratios $fS/M_\mathrm{Pl}$ and $\cT_{(2)}/(f M_\mathrm{Pl})$, which are predicted to be $\lesssim O(1)$ by the electric and magnetic axion WGC respectively, and the ratio of the axion decay constant to the string scale, to see how an experimental measurement of $f$ might shed light on the fundamental UV cutoff in our universe.

\subsection{``Model-independent'' $B^{(2)}$ axion}

One of the original examples of an extra-dimensional axion is the so-called ``model-independent string axion,'' which is just the 10d 2-form gauge field $B^{(2)}$, viewed as a 4d 2-form gauge field that is dual to an axion $\theta$~\cite{Witten:1984dg}. (The adjective ``model-independent'' is, in light of what we now know about the string theory landscape, no longer very fitting.) In the Type I or heterotic context, in our conventions, the relevant kinetic terms in the 10d effective action for $B^{(2)}$ and the $\mathrm{Spin}(32)/\ZZ_2$ gauge fields look like\footnote{My conventions follow~\cite{Svrcek:2006yi}, except that their field strengths have integer fluxes whereas mine have $2\cpi$ times integer fluxes.} 
\begin{equation}
\int  \left(-\frac{1}{2\cpi g_s^2 \ell_s^4} \frac{1}{2} H^{(3)} \wedge \star H^{(3)} - \frac{1}{4\cpi g_s^2 \ell_s^6} \frac{1}{2} \mathrm{tr}(F \wedge \star F)\right),
\end{equation}
where $H^{(3)} = \dif B^{(2)}$, $\ell_s \equiv 2\cpi \sqrt{\alpha'}$ is the string length scale, and $g_s$ is the string coupling constant. The axion couples to the gauge fields due to the modified Bianchi identity
\begin{equation}
\frac{1}{2\cpi} \dif H^{(3)} = -\frac{1}{16\cpi^2} \mathrm{tr}(F \wedge F) + \cdots,
\end{equation}
which is well-known from the Green-Schwarz mechanism~\cite{Green:1984sg}.

To see that this is a theory of an extra-dimensional axion in our sense, notice that since $\theta$ is dual to $B^{(2)}$ in four dimensions, it originates from dimensional reduction of the six-form gauge field $B^{(6)}$ that is dual to $B^{(2)}$ in 10d. The kinetic term for that field is:
\begin{equation}
-\int \frac{g_s^2 \ell_s^4}{2\cpi} \frac{1}{2} H^{(7)} \wedge \star H^{(7)}.
\end{equation}
After dualizing, the axion coupling is more straightforward, because the modified Bianchi identity becomes a standard Chern-Simons term
\begin{equation}
\frac{1}{16\cpi^2} \int B^{(6)} \wedge \mathrm{tr}(F \wedge F).
\end{equation}
Thus, everything proceeds as in \S\ref{sec:xdaxions}, but this case is relatively straightforward because all of the integrals are over {\em all} six of the extra dimensions.

We assume there is no warping and that the dilaton is constant over the internal dimensions. The axion comes from reducing $B^{(6)}$ along the 6-cycle that is simply the entire manifold $Y$. The volume form $\star_Y 1$ is harmonic, because $\dif(1) = 0$ and $\dif(\star_Y 1) = 0$ as $\star_Y 1$ is a top form. Hence the harmonic form normalized to have integral $1$ over $Y$ is $\omega^{(6)} = (\star_Y 1)/\mathrm{Vol}(Y)$. The axion string is a fundamental string, not wrapped on any internal dimensions. Hence we have:
\begin{align}
\cT_{(2)} &= \frac{2\cpi}{\ell_s^2}, &
S_\mathrm{inst} &= \frac{8\cpi^2}{g_\mathrm{YM}^2} = \frac{2\cpi \mathrm{Vol}(Y)}{g_s^2 \ell_s^6}, &
f^2 &= \frac{g_s^2 \ell_s^4}{2\cpi}\int_Y \frac{\star_Y 1 \wedge 1}{\mathrm{Vol}(Y)^2} = \frac{g_s^2 \ell_s^4}{2\cpi \mathrm{Vol}(Y)}.
\end{align}
Putting these together, we find that $\cT_{(2)} = 2\cpi \, S_\mathrm{inst}\, f^2$. This is our first example where this scaling relation holds precisely.

We can also compare these scales to expectations from the electric and magnetic axion WGC. For this we need the Planck scale, given by
\begin{equation}
M_\mathrm{Pl}^2 = 4\cpi \frac{\mathrm{Vol}(Y)}{g_s^2 \ell_s^8}.
\end{equation}
Thus we see that in this example
\begin{equation}
f S_\mathrm{inst} = \frac{\sqrt{2\cpi \mathrm{Vol}(Y)}}{g_s \ell_s^4} = \frac{1}{\sqrt{2}} M_\mathrm{Pl},
\end{equation}
parametrically saturating the naive axion WGC. We also have
\begin{equation}
\frac{1}{2\cpi f} \cT_{(2)} = \frac{1}{\sqrt{2}} M_\mathrm{Pl},
\end{equation}
parametrically saturating the naive {\em magnetic} axion WGC. As discussed in \S\ref{subsec:magneticelectric}, we believe it is not an accident that both the naive electric and magnetic axion WGCs are parametrically saturated at the same time. In our next examples, we will see that both fail to be saturated, by a common factor.

Finally, we would like to compare the axion decay constant $f$ to the string mass scale 
\begin{equation}
M_s \equiv \frac{2\cpi}{\ell_s},
\end{equation}
which is related to the Planck scale and the dimensionless volume ${\cal V} = \mathrm{Vol}(Y)/\ell_s^6$ via
\begin{equation}
M_s^2 = \frac{\cpi g_s^2}{\cal V} M_\mathrm{Pl}^2.
\end{equation}
In terms of the quantity $\alpha_\textsc{YM} = g_\textsc{YM}^2/(4\cpi)$, we have 
\begin{equation}
f = \frac{\sqrt{\alpha_\textsc{YM}}}{(2\cpi)^{3/2}} M_s.
\end{equation}
Quantitatively, if we take as a reference the GUT value $\alpha_\textsc{YM} \approx 1/25$, we have
\begin{equation}
f \approx 0.013 M_s,
\end{equation}
and the string scale is about a factor of 80 above the axion decay constant. (The definition of $M_s$ varies in the literature; in~\cite{Svrcek:2006yi} it is taken to be $1/\ell_s$. The masses of string oscillator states are closer to $2\cpi/\ell_s$, hence our choice.)

\subsection{$C^{(4)}$ axions in Type IIB string theory}
\label{subsec:Cfouraxions}

Many phenomenological investigations of Standard Model-like physics, including axions, have focused on Type IIB string theory compactified on a 6d Calabi-Yau orientifold $Y$, with Standard Model gauge fields living on D7 branes that wrap 4-cycles, which are also known as divisors (complex codimension-1 submanifolds). In this case, axions arise from the bulk gauge field $C^{(4)}$ dimensionally reduced on the 4-cycles. Axion strings are D3 branes wrapped on dual 2-cycles. Euclidean D3 branes wrapped on the 4-cycles give rise to instantons.

In this context, the geometry simplifies significantly because the volumes of interest are all calibrated by a K\"ahler 2-form $J$. Integrals of $J$, $J \wedge J$, and $J \wedge J \wedge J$ measure the volumes of 2-cycles, 4-cycles, and the entire manifold $Y$, respectively. This structure leads to a precise equation that generalizes our basic scaling relation~\eqref{eq:TSfsq} to the full set of $C^{(4)}$ axions.

For the purposes of this discussion, I will treat the dilaton field as a constant over the extra dimensions, so that I only have to keep track of powers of $g_s$. This is never truly valid in the presence of D7 branes, where a proper treatment would take the varying axiodilaton field into account using the F-theory description. However, it is sufficient for a discussion of the scaling relations of interest.

\subsubsection{Elements of Calabi-Yau geometry}

Here we collect some key formulas in Calabi-Yau geometry. Readers can find more detailed pedagogical introductions in, for example,~\cite{Becker:2006dvp, Denef:2008wq, Ibanez:2012zz}. A useful reference focused specifically on axions is~\cite{Demirtas:2018akl}.

The K\"ahler form of a Calabi-Yau threefold $Y$ can be expressed as $J = \sum t_i D^i$, where $D^i$ is a 2-form Poincar\'e dual to a divisor $\Delta_i$.\footnote{A divisor is a formal integer linear combination of holomorphic hypersurfaces (which are 4-manifolds) in $Y$.} The coefficients $t_i$ are known as K\"ahler parameters. The volume of $Y$ is a cubic function of the $t_i$,
\begin{equation} \label{eq:volumettt}
{\cal V} = \frac{1}{6} n^{ijk} t_i t_j t_k,
\end{equation}
where the $n^{ijk} \in \ZZ$ are integer intersection numbers. The 4-cycle volumes $\tau^i$ of these divisors are related to the K\"ahler parameters according to
\begin{equation}\label{eq:taudef}
\tau^i = \frac{\partial {\cal V}}{\partial t_i}.
\end{equation}
In the literature, one finds a convention in which the $t$ variables have raised indices and the $\tau$ variables have lowered indices. Here I am doing precisely the opposite, to agree with our earlier convention of raised indices on $\theta^i$ (which partners with $\tau^i$ in a chiral superfield). I hope that this does not cause confusion.

The K\"ahler parameters can be loosely thought of as measuring 2-cycle volumes, but this isn't precisely right; for instance, they are sometimes negative. A more correct statement is the following: there is an allowed set of values of the $t_i$, known as the {\em K\"ahler cone}, for which one can integrate powers of $J$ over holomorphic subvarieties to obtain positive volumes. Within the K\"ahler cone, the $\tau^i$ are positive. A 2-cycle $\Gamma$ can be characterized by its intersection numbers $w^i$ with the divisors $\Delta_i$. Certain 2-cycles are holomorphic and have volumes determined by integrating $J$ over $\Gamma$; in this case, the volume is given by $w^i t_i$. The set of $\Gamma$ for which this holds is known as the {\em Mori cone}. 

For now we will treat $t$, $\tau$, and ${\cal V}$ as dimensionless numbers, and restore units when we translate our mathematical result into physics below. Viewing the $t_i$ and thus also ${\cal V}$ as implicit functions of the $\tau^i$, the K\"ahler potential $k$ is defined as
\begin{equation} \label{eq:kahlerpotential1}
k(\{\tau^i\}) = -2 \log {\cal V}.
\end{equation}
From these definitions, it follows that
\begin{equation} \label{eq:ttaurelation}
t_i = \frac{\partial^2 k}{\partial \tau^i \partial \tau^j} \tau^j {\cal V}.
\end{equation}
We give the full argument in Appendix~\ref{app:cyderivation}.

\subsubsection{Translation to physical quantities}

Now, we translate the dimensionless volumes to physical units and explain how they relate to gauge couplings, axion decay constants, and axion string tensions. Dimensionless volumes are taken to be measured in units of $\ell_s$, so that ${\cal V} = \mathrm{Vol}(Y)/\ell_s^6$. Thus, we have
\begin{equation} \label{eq:MplVrelation}
M_\mathrm{Pl}^2 = \frac{4\cpi}{g_s^2 \ell_s^2} {\cal V}.
\end{equation}
(In our convention, Einstein and string frame metrics are related by $g_{\mu \nu,E} = \exp(-(\phi - \phi_0)/2) g_{\mu \nu,S}$ where $\exp(\phi_0) = g_s$. Thus, to the extent that the dilaton is constant, volumes are the same in both frames.) 

From the DBI action for a U(1) gauge field, with fluxes normalized to be in $2\cpi \ZZ$, on a D7 brane wrapped on a 4-cycle $\Sigma^{(4)}_i$, we read off the 4d U(1) gauge coupling $e$ as
\begin{equation}
\frac{1}{e^2} = \frac{1}{2\cpi g_s} \frac{\mathrm{Vol}(\Sigma^{(4)}_i)}{\ell_s^4} = \frac{1}{2\cpi g_s} \tau^i. 
\end{equation}
From this we infer that the 4-cycle volume $\tau^i$ in units of $\ell_s^4$ is packaged with an axion $\theta^i \cong \theta^i+2 \cpi$ arising from $C^{(4)}$, in a complex scalar field $T^i$ that is a component of an ${\cal N} = 1$ chiral multiplet $\bm{T}^i$:
\begin{equation} \label{eq:chiralT}
\bm{T}^i = \frac{1}{g_s} \tau^i - \iu \frac{\theta^i}{2\cpi} + \cdots.
\end{equation}
The factor of $1/g_s$ is sometimes absorbed in the definition of $\tau$, or into a different convention for Einstein-frame volumes relative to string-frame volumes. In the nonabelian case, rather than $\tau/g_s = 2\cpi/e^2$ we have $\tau/g_s = 4\cpi/g^2$, for the conventional normalization of the gauge coupling. In any case, the instanton action for gauge fields on a 4-cycle (or a Euclidean D3-brane on the same cycle) is 
\begin{equation}\label{eq:Staurelation}
S^i = 2\cpi \tau^i/g_s.
\end{equation}

The shift symmetry of $\theta^i$ ensures that the K\"ahler potential is a function of $\bm{T}^i + \bm{T}^{i\dagger}$, up to nonperturbatively small corrections. The leading $\bm{T}^i$-dependence is captured by the function $k$ introduced above,
\begin{equation} \label{eq:kahlerpotential2}
K(\{\bm{T}^i, {\bm{T}^i}^\dagger\}) = M_\mathrm{Pl}^2 k(\{g_s(\bm{T}^i + {\bm{T}^i}^\dagger)/2\}) + \cdots.
\end{equation}
The scalar kinetic terms take the form
\begin{equation}
-\frac{\partial^2 K}{\partial T^i \partial {T^\dagger}^j} \partial_\mu T^i \partial^\mu {T^\dagger}^j = -\frac{1}{4} \frac{\partial^2 k}{\partial \tau^i \partial \tau^j} \left[\partial_\mu \tau^i \partial^\mu \tau^j + \frac{g_s^2}{4\cpi^2} \partial_\mu \theta^i \partial^\mu \theta^j\right],
\end{equation}
from which we read off that the kinetic matrix of the axions in the normalization of~\eqref{eq:axionkinetic} is
\begin{equation} \label{eq:kappaijequation}
\kappa_{ij} = \frac{g_s^2}{8\cpi^2} M_\mathrm{Pl}^2 \frac{\partial^2 k}{\partial \tau^i \partial \tau^j}.
\end{equation}

Finally, a D3 brane has tension $\cT_{(4)} = \frac{2\cpi}{g_s\ell_s^4}$. We can obtain an axion string from wrapping a D3 brane on a 2-cycle $\Gamma$ that has intersection numbers $w^i$ with the divisors $\Delta_i$. Physically, $w^i$ represents the winding number of the axion $\theta^i$ around the string. When $\Gamma$ is in the Mori cone, this axion string has tension 
\begin{equation} \label{eq:stringtrelation}
\cT_{(2)}^{\{w\}} = \frac{2\cpi}{g_s \ell_s^2} w^i t_i.
\end{equation}
Substituting~\eqref{eq:Staurelation},~\eqref{eq:kappaijequation},~\eqref{eq:stringtrelation}, and~\eqref{eq:MplVrelation} into~\eqref{eq:ttaurelation}, we find:
\begin{equation} \label{eq:multistringtension}
\cT_{(2)}^{\{w\}}  = 2\cpi w^i \kappa_{ij} S^j.
\end{equation}
This is a precise multi-axion generalization of~\eqref{eq:TSfsq}.


\subsubsection{First example: 3-axion, fibered ``Swiss cheese'' Calabi-Yau}
\label{subsubsec:swisscheese}

Now, we consider a concrete example where there $Y$ has a large overall volume but also contains a much smaller cycle. For concreteness, we consider the Calabi-Yau manifold described in \S3.1 of~\cite{Cicoli:2016xae} (see also~\cite{Cicoli:2008gp}), which has 3 axions arising from $C^{(4)}$ reduced on 3 different 4-cycles. The volume takes the form 
\begin{equation}
{\cal V} = 9 t_1 t_2^2 + \frac{1}{6} t_3^3 = \frac{1}{6} \sqrt{\tau^1} \tau^2 - \frac{\sqrt{2}}{3} \smash{\left(\tau^3\right)}^{3/2},
\end{equation}
with $t_1 = \frac{1}{6} \frac{\tau^2}{\sqrt{\tau^1}}$, $t_2 = \frac{1}{3}\sqrt{\tau^1}$ , and $t_3 = -\sqrt{2 \tau^3}$.\footnote{In the notation of~\cite{Cicoli:2016xae}, our $(t_1, t_2, t_3)$ are their $(t_b, t_f, t_s)$ and our $(\tau^1, \tau^2, \tau^3)$ are their $(\tau_f, \tau_b, \tau_s)$. Their subscripts are $b$ for big, $f$ for fiber, and $s$ for small. Their choice uses the same subscripts for $t_f$ and $\tau_f$, because both are related to the fiber size, but they are not linked by~\eqref{eq:taudef}. We take our numerical labels as in~\eqref{eq:taudef}, so the fiber corresponds to $t_2$ but $\tau^1$.} Within the K\"ahler cone, $t_1, t_2 > 0$ and $t_3 < 0$. This structure reflects a manifold which, on large scales, is a four-dimensional K3 manifold of volume $\tau^1 \sim \smash{(t_2)}^2$ fibered over a two-dimensional $\mathbb{P}^1$ base of volume $t_1$, and which contains a small, self-intersecting divisor of volume $\tau^3$. We work in the limit ${\cal V} \gg \smash{\left(\tau^3\right)}^{3/2}$, where there is a small hole in a much larger manifold. This is a prototype for a much larger family of examples which might have multiple such small holes, some of which could play a role in stabilizing the overall volume modulus~\cite{Balasubramanian:2005zx} while others could host the Standard Model gauge fields on wrapped D7 branes. We will consider a similar but more subtle example in \S\ref{subsubsec:swisscheese1}.

In this example, the axion kinetic terms are given by
\begin{equation}
-\frac{1}{2} \left(\frac{g_s^2}{8\cpi^2} M_\mathrm{Pl}^2\right) \left[\frac{1}{(\tau^1)^2} \rmd \theta^1 \wedge \star \rmd \theta^1+\frac{2}{(\tau^2)^2} \rmd \theta^2 \wedge \star \rmd \theta^2  + \frac{1}{\sqrt{2 \tau^3} {\cal V}} \rmd \theta^3 \wedge \star \rmd \theta^3 + \cdots \right].
\end{equation}
The $\cdots$ represents subleading terms, including kinetic mixing among the axions as well as volume-suppressed contributions to the kinetic terms displayed. Although the mixing contributions are volume suppressed, they are not necessarily negligible in the relation~\eqref{eq:multistringtension}, because some of the $S^j$ are enhanced by large volumes. 

In the limit that $\tau^3$ goes to zero, the volume factors into the product of $\sqrt{\tau^1}$ and $\tau^2$, and so the moduli $\tau^1$ and $\tau^2$ each have logarithmic kinetic terms of the form familiar in infinite distance limits of moduli space (see, e.g.,~\cite{Ooguri:2006in}). They mix only through subleading terms involving $\tau^3$. Correspondingly, for the axions $\theta^1$ and $\theta^2$, the single-axion relationship~\eqref{eq:TSfsq} holds up to volume-suppressed terms. Furthermore, in both cases the naive electric and magnetic axion WGC relations for $f S_\mathrm{inst}$ and $\cT_{(2)}/(2\cpi f)$ hold: specifically,
\begin{alignat}{3}
& \theta^1: \qquad && f_1 S^{(1)} = \frac{1}{\sqrt{2}} M_\mathrm{Pl}, \qquad&& \frac{1}{2\cpi f_2} \cT_{(2)}^{\{1,0,0\}} = \frac{1}{\sqrt{2}} M_\mathrm{Pl}; \nonumber \\
& \theta^2: \qquad && f_2 S^{(2)} = M_\mathrm{Pl}, \qquad && \frac{1}{2\cpi f_2}\cT_{(2)}^{\{0,1,0\}}  = M_\mathrm{Pl}.
\end{alignat}
This is no surprise; we expect all of these relationships to come as a package for axions associated with moduli that have infinite distance limits. Notice, also, that in both cases the coefficient in the electric and magnetic WGC relationships is the same, along the lines we argued in \S\ref{subsec:magneticelectric}. Here, and elsewhere, I have put parentheses around the numerical superscript on the action in places where it might otherwise be confused with an exponent. 

There is no infinite-distance limit in moduli space associated with $\tau^3$ alone. Because it contributes negatively to the overall volume, we cannot make $\tau^3$ arbitrarily large while holding $\tau^1$ and $\tau^2$ fixed. Thus, the kinetic term for $\tau^3$ does not have the universal logarithmic form expected for moduli with asymptotic limits; in fact, $\tau^3 \to 0$ is a finite-distance limit. The kinetic term for $\theta^3$, proportional to $1/\sqrt{\tau^3}$, is of the power law form discussed in~\S\ref{subsubsec:powerlawex} with $\gamma = -1/4$. This is the case $\gamma > -1/2$ where we found no physical Euclidean instanton solutions. However, the analysis of~\S\ref{subsubsec:powerlawex} breaks down here because the $1/\sqrt{\tau^3}$ dependence is only an approximation, valid when $\tau^3 \ll \tau^{1,2}$.  Relatedly, the kinetic term for $\tau^3$ is highly dependent on the values of $\tau^1$ and $\tau^2$, and mixing terms are important in the relationship~\eqref{eq:ttaurelation}. Specifically, we have
\begin{equation}
\frac{\partial^2 k}{\partial \tau^3 \partial \tau^1} \tau^1 {\cal V} \approx -\frac{1}{\sqrt{2}} \sqrt{\tau^3}; \quad \frac{\partial^2 k}{\partial \tau^3 \partial \tau^2} \tau^2 {\cal V}\approx -\sqrt{2\tau^3}; \quad \frac{\partial^2 k}{\partial \tau^3 \partial \tau^3} \tau^3 {\cal V}\approx \frac{1}{\sqrt{2}} \sqrt{\tau^3}.
\end{equation}
Here $\approx$ indicates that we have dropped volume-suppressed terms. Summing these up, we obtain $-\sqrt{2 \tau^3} = t_3$, as promised by~\eqref{eq:ttaurelation}. Interestingly, if we kept only the diagonal term, we would obtain parametrically the same answer, but off by a factor of $-1/2$. Thus, if we focus on the axion $\theta^3$ alone, the holomorphic axion string of minimal winding has $w^3 = -1$ and hence
\begin{equation}
\cT_{(2)}^{\{0,0,-1\}} \approx 4\cpi S^{(3)} f_3^2.
\end{equation}
Thus, the single-field parametric relation~\eqref{eq:TSfsq} is still a good guide to the scale of the axion string tension, even though getting the order-one coefficient in the relationship right requires us to use the multi-field formula~\eqref{eq:multistringtension}. Furthermore, in this case the naive electric and magnetic axion WGC are both satisfied by a large parametric factor:
\begin{equation} \label{eq:theta3parametrics}
\theta^3: \qquad f_3 S^{(3)} = \frac{\smash{\left(\tau^3\right)}^{3/4}}{2^{3/4}\sqrt{\cal V}}M_\mathrm{Pl}, \qquad \frac{1}{2\cpi f_3} \cT_{(2)}^{\{0,0,-1\}} =  \frac{2^{1/4} \smash{\left(\tau^3\right)}^{3/4}}{\sqrt{\cal V}} M_\mathrm{Pl}.
\end{equation}
This is consistent with our expectation, from \S\ref{subsec:magneticelectric}, that the parametric coefficients appearing in the electric and magnetic axion WGC relationships should be the same. In this case, the quantitative match is not perfect, due to factors of $2^{1/4}$; this is again related to effects of mixing.

Finally, we evaluate the decay constant of $\theta^3$:
\begin{equation}
f_3 = \frac{g_s M_\mathrm{Pl}}{2^{7/4} \cpi {\cal V}^{1/2} \smash{\left(\tau^3\right)}^{1/4}} = \frac{1}{2^{7/4} \cpi^{3/2} \smash{\left(\tau^3\right)}^{1/4}} M_s.
\end{equation}
Thus, the axion decay constant is near the string scale. In particular, if we have in mind an axion coupled to the Standard Model and set $\tau^3 = g_s \alpha_\textsc{GUT}^{-1}$ with the conventional GUT value $\alpha_\textsc{GUT} \approx 1/25$, then numerically we have
\begin{equation}
f_3 \approx 0.024 g_s^{-1/4} M_s.
\end{equation}
The $g_s$ dependence is very mild (and $g_s$ should not be too small, or we would not be able to achieve large enough Standard Model couplings while keeping volumes large enough to trust our calculations), so we conclude that the string scale in such a model should be around 20 to 40 times the axion decay constant. The other decay constants, $f_1$ and $f_2$, have a less straightforward relationship to the string scale. When the overall volume is approximately isotropic, in the sense that $\tau^1 \sim \tau^2$, these decay constants are both near the Kaluza-Klein scale $M_\textsc{KK} \sim M_\mathrm{Pl}/{\cal V}^{2/3}$. Taking $\tau^1 \gg \tau^2$ raises $M_s$ relative to $f_1$, and taking $\tau^2 \gg \tau^1$ raises $M_s$ relative to $f_2$. In any case, if we focus on $f_i$ while fixing $\tau^i = g_s \alpha_\textsc{GUT}^{-1}$, the only way to achieve $M_s \gg 10^2 f_i$ is to make some other $\tau^j$ small in string units, at which point the calculation is not under control.

\subsubsection{Second example: 2-axion ``Swiss cheese'' Calabi-Yau}
\label{subsubsec:swisscheese1}

Next, we consider a classic example which has only two axions instead of 3, and a similar structure to the previous example. Despite initial appearances, this example is actually less straightforward, because the axion string of minimal tension turns out to not be the axion string of minimal winding. This example is a Calabi-Yau with $h^{1,1} = 2$, the anticanonical hypersurface in the weighted projective space $\mathbb{P}_{[1,1,1,6,9]}$. This is a well-studied CY manifold often discussed in the context of moduli stabilization scenarios~\cite{Candelas:1994hw, Denef:2004dm, Balasubramanian:2005zx}. We will work in the basis chosen in section 4.1 of the CYTools manual, which discusses many properties of this manifold very concretely~\cite{Demirtas:2022hqf}.

We denote the Poincar\'e duals to the two divisors in the chosen basis as $D^1$ and $D^2$, and parametrize the K\"ahler form as $J = t_1 D^1 + t_2 D^2$. The volume is
\begin{equation}
{\cal V} = \frac{1}{18} \left[t_1^3 - \left(t_1 - 3 t_2\right)^3\right].
\end{equation}
The K\"ahler cone, i.e., the range of parameters for which we can compute positive volumes by integrating powers of $J$ over subvarieties, is given by imposing the conditions
\begin{equation} \label{eq:kahlercone}
t_1 - 3 t_2 \geq 0\quad \text{and} \quad t_2 \geq 0.
\end{equation}
Thus we can think of $t_1$ as parametrizing the overall volume of the Calabi-Yau, and $t_1 - 3 t_2$ as parametrizing the size of a hole in the Calabi-Yau.

The divisor volumes are computed as
\begin{align}
\tau^1 = \frac{\partial {\cal V}}{\partial t_1} &= \frac{1}{2}t_2 \left(2 t_1 - 3 t_2\right), \nonumber \\
\tau^2 = \frac{\partial {\cal V}}{\partial t_2} &= \frac{1}{2} \left(t_1 - 3 t_2\right)^2.
\end{align}
We see that $D^2$ is the small divisor, with $\tau^2 \to 0$ at the edge of the K\"ahler cone where $t_1 \to 3 t_2$. We could carry out a $\mathrm{GL}(2,\ZZ)$ transformation of our basis of divisors to make one of the K\"ahler parameters $t_i$ vanish as $t_1 \to 3 t_2$, but then the conjugate $\tau^i$ would not vanish in this limit.\footnote{In the literature, one sometimes sees the formula written as ${\cal V} \sim \tau_b^{3/2} - \tau_s^{3/2}$ for a big volume $\tau_b$ and small volume $\tau_s$, with $\tau_b \sim t_b^2$ and $\tau_s \sim t_s^2$ in terms of big and small 2-cycle volumes $t_b$ and $t_s$. Such a form can only be obtained through a non-integral basis choice of divisors, which obscures the possible allowed charges of instantons and axion strings.} The overall volume can also be expressed as
\begin{equation}
{\cal V} = \frac{\sqrt{2}}{9} \left[\left(3 \tau^1 + \tau^2\right)^{3/2} - \left(\tau^2\right)^{3/2}\right],
\end{equation}
in terms of the divisor volumes.

An axion string is obtained by wrapping a D3 brane on a curve $C$ that has intersection numbers $w^i$ with the divisors $D^i$. The volume of such a curve is
\begin{equation}
\mathrm{Vol}(C) = w^i t^i.
\end{equation}
This formula is reliable and computes positive volumes for curves in the Mori cone, which obey the inequalities
\begin{equation} \label{eq:moricone}
w^1 \geq 0\quad \text{and} \quad  w^2 + 3 w^1 \geq 0.
\end{equation}
In particular, there is a small volume curve with $w^1 = 1$, $w^2 = -3$, and volume $t_1 - 3 t_2$, which wraps a 2-cycle in the Calabi-Yau that shrinks to zero size at the edge of the K\"ahler cone.

The axion kinetic terms are given by
\begin{equation}
-\frac{1}{2} \left(\frac{g_s^2}{8\cpi^2} M_\mathrm{Pl}^2\right) \left[\frac{3}{(\tau^1)^2} \rmd \theta^1 \wedge \star \rmd \theta^1+\frac{1}{3\sqrt{2 \tau^2}{\cal V}} \rmd \theta^2 \wedge \star \rmd \theta^2 + \frac{2}{(\tau^1)^2} \rmd \theta^1 \wedge \star \rmd \theta^2 + \cdots \right].
\end{equation}
The third term is the leading kinetic mixing between the two axions; the $\cdots$ represents contributions suppressed by powers of the small ratio $\sqrt{\tau^2/\tau^1}$. 

If we focus on one axion at a time, inspecting the corresponding instanton action and the tension of a string of minimal winding, we have (up to subleading terms):
\begin{alignat}{3}
& \theta^1: \qquad && f_1 S^{(1)} = \sqrt{\frac{3}{2}} M_\mathrm{Pl}, \qquad&& \frac{1}{2\cpi f_1} \cT_{(2)}^{\{1,0\}} =\sqrt{\frac{3}{2}} M_\mathrm{Pl}; \nonumber \\
& \theta^2: \qquad && f_2 S^{(2)} = \frac{\smash{\left(\tau^2\right)}^{3/4}}{2^{3/4}\sqrt{3 {\cal V}}} M_\mathrm{Pl}, \qquad && \frac{1}{2\cpi f_2} \cT_{(2)}^{\{0,1\}} = \sqrt{\frac{1}{6}} M_\mathrm{Pl}.
\end{alignat}
Here we see a breakdown of the pattern we have seen before. For the axion corresponding to the large volume modulus, the naive electric and magnetic WGC hold with the same coefficient. For the axion $\theta^2$ corresponding to the small volume modulus, the instanton action is set by the small quantity $\tau^2$, but the corresponding axion string with $w^1 = 0, w^2 = 1$ has a tension set by $t_2$, which is not small. The naive magnetic WGC estimate holds, but $f S_\mathrm{inst}$ is below the naive electric WGC estimate by a large factor. Our basic scaling estimate~\eqref{eq:TSfsq} breaks down badly in this case. On the other hand, the axion string with $w^1 = 1$ and $w^2 = -3$ has tension
\begin{equation}
\frac{1}{2\cpi f_2} \cT_{(2)}^{\{1,-3\}} = \frac{3^{1/2} 2^{1/4} \smash{\left(\tau^2\right)}^{3/4}}{\sqrt{\cal V}} M_\mathrm{Pl}.
\end{equation}
If we take this minimal tension string with nontrivial $\theta^1$ winding to be the string whose tension should appear in the estimate~\eqref{eq:TSfsq}, then the pattern holds. We have
\begin{equation}
\frac{1}{2\cpi f_2^2 S^{(2)}}   \cT_{(2)}^{\{1,-3\}} = 6.
\end{equation}

Finally, we evaluate the decay constant of $\theta^2$:
\begin{equation}
f_2 = \frac{g_s M_\mathrm{Pl}}{3^{1/2} 2^{7/4} \cpi {\cal V}^{1/2} \smash{\left(\tau^2\right)}^{1/4}} = \frac{1}{3^{1/2} 2^{7/4} \cpi^{3/2} \smash{\left(\tau^2\right)}^{1/4}} M_s \approx 0.014 g_s^{-1/4} M_s.
\end{equation}
As in our other examples, we see that the fundamental string scale can be at most about two orders of magnitude above the axion decay constant.

\subsubsection{Comments on the Calabi-Yau examples}

In the two examples we have considered, the axions that have a decay constant parametrically below the naive WGC bound, as in~\eqref{eq:theta3parametrics}, are paired in a chiral superfield~\eqref{eq:chiralT} with a modulus $\tau$ describing the volume of a collapsing cycle within the larger CY manifold. The small-volume limit $\tau \to 0$ lies at finite distance in field space, but is at the boundary of the K\"ahler cone. There are several different ways that a boundary of the K\"ahler cone can arise in moduli space~\cite{Witten:1996qb, Mayer:2003zk, Brodie:2021nit}. These include: flop walls, at which the volume of a curve (2-cycle) goes to zero while divisor volumes and the overall volume of $Y$ remain finite; Zariski walls, at which the volume of a divisor goes to zero while the overall volume remains finite; and effective cone walls, where the overall volume of $Y$ goes to zero. Zariski walls come in two types, depending on whether the divisor collapses to a finite-volume curve or to a point. Our examples are Zariski walls of the second type: a divisor collapses to a point, as measured by $\tau$, and it intersects itself along a curve that {\em also} collapses to a point, as measured by a linear combination of K\"ahler parameters $t \sim \sqrt{\tau}$. This scaling explains why our examples obey the basic relation~\eqref{eq:TSfsq}: the axion string tension is proportional to $t$, the instanton action to $\tau \sim t^2$, and the square of the decay constant to $1/\sqrt{\tau} \sim 1/t$. All of these quantities are determined locally, on the small cycle, independent of the large-scale geometry of the Calabi-Yau. Notice that to make this scaling argument, we do not need to be near the boundary of the K\"ahler cone in the absolute sense, with $\tau \ll 1$; indeed, we want $\tau = \alpha^{-1} \gg 1$, to reproduce the small Standard Model gauge couplings. However, we are working near the boundary of the K\"ahler cone in the {\em relative} sense that $\tau$ is small relative to the size of the overall Calabi-Yau volume. The twist that we saw in~\eqref{subsubsec:swisscheese1}, where the axion string corresponding to the small curve has nontrivial winding not only for the axion reduced on the small divisor but for another axion, depends on how the local physics of the collapsing cycle is embedded in the larger Calabi-Yau, via the intersection numbers between small and large divisors.

It would be interesting to extend our study to a larger range of Calabi-Yau examples. One direction to pursue would be a detailed understanding of simple examples with a small number of moduli, near different types of boundaries of the K\"ahler cone. Another would be a statistical survey of a much larger family of Calabi-Yau manifolds, along the lines of~\cite{Demirtas:2018akl, Halverson:2019cmy, Mehta:2021pwf, Broeckel:2021dpz, Demirtas:2021gsq, Gendler:2023kjt}. It would also be interesting to explore Type IIA examples, where axions arise from gauge fields integrated over odd-dimensional cycles.

\subsection{Type IIB axion $C_0$}
\label{subsec:Czero}

Finally, we consider an interesting edge case: the 4d axion that simply coincides with the 10d axion $C_0$ in Type IIB string theory rather than arising by integrating a higher-form field over a cycle. In this case, the instantons already exist in 10d. They are simply D$(-1)$-branes or D-instantons, which have action
\begin{equation}
S_\mathrm{inst} = \frac{2\cpi}{g_s}. 
\end{equation}
Axion strings come from D7 branes wrapped on all six internal dimensions, with 4d tension
\begin{equation}
\cT_{(2)} = \cT_{(D7)} \mathrm{Vol}(Y) = \frac{2\cpi {\cal V}}{g_s \ell_s^2}.
\end{equation}
Finally, dimensional reduction of the 10d kinetic term for $C_0$ gives
\begin{equation}
f^2 = \frac{1}{2\cpi \ell_s^8} \mathrm{Vol}(Y) = \frac{{\cal V}}{2\cpi \ell_s^2}.
\end{equation}
We see, once again, that in this example $\cT_{(2)} = 2\cpi S_\mathrm{inst} f^2$.

The naive axion electric and magnetic WGC formulas also hold in this case:
\begin{equation}
fS_\mathrm{inst} = \frac{1}{2\cpi f} \cT_{(2)} = \frac{1}{\sqrt{2}} M_\mathrm{Pl}.
\end{equation}
Finally, as in the examples above, we can compare the axion decay constant to the string scale. We have
\begin{equation}
f = \frac{\cal V}{(2\cpi)^{3/2}} M_s.
\end{equation}
For our calculations to be under control we should have ${\cal V} \gg 1$, so in this case the string scale is expected to be at or below the scale $f$.

This case is less directly phenomenologically relevant than the other examples above. The axion $C_0$ couples to gauge fields that live on D3 branes. However, D3s in the bulk lack chiral matter. To obtain a Standard Model-like theory with chiral matter, we can place D3 branes at a singularity, but then we must consider additional moduli localized at the singularity to obtain a complete picture of the physics.

\section{Comments on chiral fermions and axions}
\label{sec:chiralfermions}

In this section, we comment on two aspects of axion physics related to chiral fermions, based on expectations from string theory. First, we argue that axions are unlikely to acquire a large tree-level mass in theories with chiral charged fermions. Second, we comment on the expected size of derivative couplings of an extra-dimensional axion to fermions, which differs from the size expected in 4d models of either KSVZ or DFSZ type. The remarks in \S\ref{subsubsec:monodromyagain} are original, whereas most of the remainder of this section gives my viewpoint on results that have been discussed elsewhere.

\subsection{Tree-level masses, revisited}
\label{subsec:treelevelagain}

In \S\ref{subsec:treelevelmass} we briefly discussed two possible tree-level masses for axions: a $\theta F^{(4)}$ term or monodromy mass, and a Stueckelberg mass $|\rmd \theta - q A|^2$ in which the axion is eaten by an ordinary gauge field. We will now return to these two possibilities in more specific contexts. In the monodromy case, we will argue in specific examples that a $\theta F^{(4)}$ mass term cannot be turned on for an axion that interacts with a chiral gauge theory like the Standard Model. In the Stueckelberg case, we point out that typically the large axion mass arises in the context of the 4d Green-Schwarz mechanism, where there are two axions. One linear combination is eaten, but another remains light. These considerations strengthen the argument for the existence of a light axion coupled to the Standard Model in the extra-dimensional scenario.

\subsubsection{Monodromy mass versus chiral fermions}
\label{subsubsec:monodromyagain}

In Type IIA string theory, a Standard Model-like theory with chiral fermions can be constructed on intersecting D6 branes (see, e.g.,~\cite{Blumenhagen:2005mu,Cvetic:2011vz} for reviews). The D6 branes are wrapped on 3-cycles to produce a 4d gauge theory. The RR form $C^{(3)}$ reduced on a 3-cycle becomes a 4d axion coupling to the gauge fields on branes wrapping that cycle, and axion strings are D4 branes wrapped on an intersecting 3-cycle.

In this particular framework, at first glance the axion quality problem appears to be endangered by the possibility of a tree-level monodromy mass.\footnote{This was pointed out in a footnote in section 7.1 of~\cite{Heidenreich:2020pkc}.} The danger is a 10d Chern-Simons coupling of the form
\begin{equation} \label{eq:10dCS}
\frac{1}{8\cpi^2} \int C^{(3)} \wedge \rmd C^{(3)} \wedge H^{(3)},
\end{equation}
where $H^{(3)} = \rmd B^{(2)}$ is the field strength of the $B$-field under which the fundamental string is charged. If we integrate the first $C^{(3)}$ over a 3-cycle $\alpha$ to obtain a 4d axion $\theta$, and we turn on a flux $\frac{1}{2\cpi} \int_\beta H^{(3)} = n \neq 0$ over an intersecting 3-cycle $\beta$, we obtain an effective 4d monodromy mass $\frac{n}{2\cpi} \int \theta\, \rmd C^{(3)}$. 

However, there is an obstruction to turning on the requisite $H^{(3)}$ flux in chiral gauge theories (like the Standard Model). In this scenario, chiral charged fermions arise at the intersection of the cycles $\alpha$ and $\beta$, when additional D6 branes are placed on $\beta$. The gauge field $A^{(1)}$ living on a D-brane has a Stueckelberg coupling to the bulk field $B^{(2)}$, i.e., its kinetic term has the form
\begin{equation}
-\frac{1}{2e^2} (F^{(2)} - B^{(2)}) \wedge {\star_{\rm D6}{(F^{(2)} - B^{(2)})}}.
\end{equation}
If we dualize the field $A^{(1)}$ on a D6 brane, we obtain a magnetic dual gauge field $A_\textsc{M}^{(4)}$ with
\begin{equation}
\frac{1}{2\cpi} \rmd A_\textsc{M}^{(4)} = \frac{1}{e^2}{ \star_{\rm D6}{\left(F^{(2)}-B^{(2)}\right)}},
\end{equation}
together with a Chern-Simons term of the form
\begin{equation}
\frac{1}{2\cpi} \int A_\textsc{M}^{(4)} \wedge H^{(3)},
\end{equation}
to reproduce the physics of the Stueckelberg interaction. Consequently, if we turn on a flux $H^{(3)}$ on $\beta$ and simultaneously wrap a D6 brane on $\beta$, we obtain a tadpole sourcing the D6$_\beta$-localized field $A_\textsc{M}^{(4)}$ in the four noncompact dimensions. This is inconsistent, provided that we do not also introduce further ingredients to cancel the tadpole, but we expect that doing so would lead to a brane configuration that somehow reconnects and removes the chiral fermions from the spectrum. This is the simple version of a story that is more subtle in the presence of torsion classes~\cite{Freed:1999vc}, and is also similar to physics discussed in~\cite{Uranga:2002vk, Behrndt:2003ih, Cascales:2003zp}.

Let us be slightly more careful about the details. In particular, what if we turn on $H^{(3)}$ flux on a cycle $\beta$ that intersects $\alpha$ but we obtain chiral fermions from the intersection of $\alpha$ with a {\em different} cycle $\gamma$? We can work with a symplectic basis of 3-cycles on the Calabi-Yau, that is, a set of $\alpha_i$ and $\beta_i$ (with $i \in \{0, \ldots, h^{2,1}\}$) with intersection numbers
\begin{equation}
\alpha_i \cap \alpha_j = 0, \quad \beta_i \cap \beta_j = 0, \quad \alpha_i \cap \beta_j = \delta_{ij}.
\end{equation}
Without loss of generality, we can choose this basis so that our initial D6$_\alpha$ branes wrap an integer multiple of the cycle $\alpha_1$, and our axion $\theta^1$ is obtained by integrating $C^{(3)}$ over this cycle. Now, it is clear that in order to give $\theta^1$ a monodromy mass, we must have $\int_{\beta_1} H^{(3)} \neq 0$. Importantly, the Chern-Simons term~\eqref{eq:10dCS} in the presence of $H^{(3)}$ flux can only give mass to {\em one} linear combination of the 4d $C^{(3)}$ axions, because there is only one 4d four-form field strength, namely $\rmd C^{(3)}$ with all legs in 4d directions. If the heavy linear combination is not precisely $\theta^1$, then we can integrate out the heavy axion, and a light axion will remain that couples to the gauge fields on the D6$_{\alpha_1}$ branes. Thus, obtaining a theory with no light axion coupled to these gauge fields requires that $\int_{\alpha_i} H^{(3)} = 0$ for all $i$, and that $\int_{\beta_i} H^{(3)} = 0$ for all $i \neq 1$ (or equivalently, that $H^{(3)}$ is proportional to the Poincar\'e dual of $\alpha_1$). Now, if we wrap a D6 brane on a general linear combination $m^i \alpha_i  + n^i \beta_i$ of 3-cycles, it will intersect the original D6$_{\alpha_1}$~branes and lead to chiral charged matter only if $n^1 \neq 0$, but this is precisely what our chosen $H^{(3)}$ flux obstructs.

An even simpler example, along similar lines, is to consider gauge fields living on D3 branes in Type IIB string theory. The bulk axion field $C^{(0)}$ couples to such gauge fields, but it can receive a large tree-level mass from $H^{(3)}$ flux via the $C^{(0)} H^{(3)} \wedge \rmd C^{(6)}$ Chern-Simons term (with $C^{(6)}$ integrated over a 3-cycle to give rise to a 4d 3-form gauge field). However, as noted at the end of \S\ref{subsec:Czero}, to obtain chiral matter from D3 branes we must place them at a singularity, and then additional localized axions appear.

We see that, at least in the simplest examples one can construct, the monodromy mechanism for providing a large tree-level axion mass fails to decouple axions from Standard Model-like theories with chiral matter.

\subsubsection{Green-Schwarz mechanism and axions}
\label{subsubsec:greenschwarz}

It is quite common for axions in string theory to obtain a mass from Stueckelberg couplings of the form $|\rmd \theta - q A|^2$ to a $\Uone$ gauge field $A$. The physics of this case is richer than that of a monodromy mass. There is a $\Uone$ gauge symmetry which is exact, and spontaneously broken by the Stueckelberg mechanism. Below the scale of the gauge field mass, an exponentially good approximate $\Uone$ global symmetry survives. This behaves as a Peccei-Quinn symmetry, and if another $\Uone$-charged field obtains a vacuum expectation value, then another (uneaten) axion will result (see, e.g.,~\cite{Bonnefoy:2018hdo} for a general argument). Many models where the QCD axion is such a secondary axion, arising from the phase of a charged field rather than an extra-dimensional gauge field, have been studied in string theory~\cite{Barr:1985hk, Kim:1988dd, Svrcek:2006yi, Choi:2011xt, Cicoli:2013cha, Honecker:2013mya, Allahverdi:2014ppa, Choi:2014uaa, Buchbinder:2014qca}. A phenomenological model of extra dimensions with similar properties was discussed in~\cite{Cheng:2001ys}. 

In many examples of this type, the Stueckelberg mass is accompanied by the 4d Green-Schwarz mechanism. Specifically, we have a coupling
\begin{equation}
\int \frac{k}{8\cpi^2} \theta F \wedge F,
\end{equation}
which is not invariant under the gauge transformation $A \mapsto A + \rmd \lambda$, $\theta \mapsto \theta + q \lambda$. However, if there are also chiral fermions that are anomalous under the $\Uone$, the anomalous term arising from their transformation can cancel the term arising from the shift of $\theta$, provided that $\sum_i q_i^3 + k q = 0$. One can also cancel $\Uone\SUN^2$ anomalies with a $\theta \,\mathrm{tr}(F \wedge F)$ term, in a variant of this mechanism.

In the supersymmetric context, we expect that often a field charged under $A$ will get a VEV at a very high scale. This is because the chiral superfield $T$ containing $\theta$, which shifts under an $A$ gauge transformation, leads to an effective (field-dependent) Fayet-Iliopoulos term~\cite{Dine:1987xk, Atick:1987gy, Dine:1987gj}, with the D-term potential for charged scalars taking the form
\begin{equation} \label{eq:VDterm}
V_D = \frac{1}{2} g^2 \left(\sum_i q_i |\phi_i|^2 - \frac{q}{4\cpi} \partial_T K\right)^2.
\end{equation}
From our examples in \S\ref{subsec:Cfouraxions}, we expect that often $\partial_T K$ will be of order $M_s^2$, up to a prefactor that depends on 4d gauge couplings and order-one numbers. Thus, some of the $\phi_i$ will typically get a VEV around the same order as the decay constant $f$ of the original axion $\theta$. The gauge field $A$ will eat a linear combination of the phases of the $\phi_i$ and $\theta$, while another linear combination remains as the light axion.

As discussed in~\cite{Heidenreich:2020pkc, Aloni:2024jpb}, when there is a chiral symmetry broken only by the ABJ anomaly, we can think of this as having gauged the $(-1)$-form global instanton number, because $\mathrm{tr}(F \wedge F) \propto \rmd J_C$ becomes exact. An axion can also be thought of as gauging the $(-1)$-form instanton number. In the Green-Schwarz scenario, we have thus effectively gauged the $(-1)$-form global instanton number {\em twice}---once with chiral fermions, and once with an axion---but then we have a remaining conserved current that is a linear combination of $J_C$ and the axion shift symmetry current $\rmd{\star \theta}$. This extra current is gauged by the ordinary $\Uone$ gauge field $A$.

The Green-Schwarz scenario, in which the light axion is a linear combination of $\theta$ and the $\arg(\phi_i)$, has the potential to maintain many of the favorable aspects of extra-dimensional axions. In particular, it can provide exponentially good control of the quality problem, because $\Uone$ gauge invariance allows us to write PQ-breaking operators built out of the $\phi_i$ fields only if dressed with an appropriate power of $\E^{-T}$. If $\langle \phi_i \rangle$ is of the same order as $f$, then our qualitative conclusions about the relationship of $f$ to the axion string tension and the UV cutoff of the theory should be unchanged. Furthermore, if there is no separation of scales, then there will still be no 4d PQ-breaking phase transition. On the other hand, if one can arrange for all VEVs $\langle \phi_i \rangle$ to be much smaller than $f$, one can have a scenario where $\theta$ decouples at a high scale, and one is left with a conventional 4d model with an approximate PQ symmetry broken by VEVs of 4d fields. Such a scenario would have a 4d PQ phase transition and solitonic axion strings. Could this provide a viable, consistent model of a post-inflation axion that evades the quality/cosmology tension that afflicts most such models~\cite{Lu:2023ayc}? The first challenge is how to suppress the VEVs relative to $f$. From~\eqref{eq:VDterm}, we see that this requires making $\partial_T K$ small. This can happen naturally for small cycles in the $\langle \mathrm{Re}\,T\rangle \to 0$ limit, where they collapse into singularities~\cite{Cicoli:2012sz, Cicoli:2013cha, Allahverdi:2014ppa}. Thus, a promising setting for realizing a conventional 4d axion cosmology in string theory is with the Standard Model localized on D3 branes at a singularity. The next challenge is to see whether this scenario can address the axion quality problem, as $\E^{-T}$ factors no longer provide exponential suppression in this limit. To the best of my knowledge, this has not been addressed in detail in the literature. It would be very interesting to further develop this scenario and to understand how one would interpret a measurement of the axion decay constant in it, and whether there are clean experimental handles to distinguish the extra-dimensional axion scenario from the Green-Schwarz axion scenario. One possibility is to go beyond the axion-photon and axion-gluon couplings and measure derivative couplings of axions to fermions, as we will now discuss.

\subsection{Fermion couplings}

Experimental searches for, and constraints on, axions depend not only on the axion coupling to gauge fields but also to Standard Model fermions. For completeness, we briefly remark on the expectations for such couplings in the case of extra-dimensional axions. Our claims in this section are not new, and one can find similar discussions in, for example,~\cite{Cicoli:2012sz, Choi:2021kuy, Choi:2024ome}. We give an argument based on a supersymmetric ansatz for the saxion dependence of the K\"ahler potential for the Standard Model fields, although we expect that the resulting estimate holds even away from the supersymmetric limit, for geometric reasons.

Consider the case where we have a modulus supermultiplet $\bm{T}$ whose scalar component includes both a noncompact modulus $\tau$ and an axion $\theta \cong \theta + 2\cpi$:
\begin{equation}
\bm{T} = \tau - \iu \frac{\theta}{2\cpi} + \cdots.
\end{equation}
Let $\bm{\Psi}$ denote a chiral superfield containing a Standard Model fermion $\psi$, and suppose the kinetic term for $\bm{\Psi}$ has a power-law dependence on the modulus $\tau$:
\begin{equation}
{\cal L} \supset \int \mathrm{d}^4 \Theta \, \bm{k_\Psi} (\bm{T} + \bm{T}^\dagger)^\gamma \bm{\Psi}^\dagger \bm{\Psi},
\end{equation}
where $\gamma$ is a constant and $\bm{k_\Psi}$ in general might depend on other moduli. Some general scaling arguments for this kind of kinetic term arising from geometry, when Standard Model fermions live on submanifolds of those on which the gauge fields live, appear in~\cite{Conlon:2006tj}.

From this expression, we can extract the $\psi$ kinetic term as well as the coupling to the axion $\theta$:
\begin{equation}
{\cal L} \supset k_\Psi (2\tau)^\gamma \psi^\dagger \overline{\sigma}^\mu \partial_\mu \psi + \gamma k_\Psi (2 \tau)^{\gamma-1} \partial_\mu \theta\, \psi^\dagger \overline{\sigma}^\mu \psi. 
\end{equation} 
Defining the canonically normalized fermion $\widehat \psi = \sqrt{k_\Psi (2\tau)^\gamma} \psi$, we see that we have a derivative coupling of the axion of the form:
\begin{equation}
\frac{\gamma }{2\tau} \partial_\mu \theta \, {\widehat \psi}^\dagger \overline{\sigma}^\mu {\widehat \psi}.
\end{equation}
The gauge coupling $g$ when the holomorphic gauge kinetic function arises from $\bm{T}$ is $\frac{g^2}{4\cpi} = \frac{1}{\tau}$, and we can introduce a canonically normalized axion $a = \theta f_a$, so that this has the form
\begin{equation}
\frac{\gamma g^2}{16\cpi^2 f_a} \partial_\mu a \, {\widehat \psi}^\dagger \overline{\sigma}^\mu {\widehat \psi}.
\end{equation}
In other words, we expect derivative couplings of axions to fermions to be essentially of one-loop size, with a weight factor $\gamma$ that could be model-dependent. 

This is an interesting result, because in 4d models of DFSZ type~\cite{Zhitnitsky:1980tq,Dine:1981rt} we can find fermion couplings that are simply an O(1) number times $1/f_a$, whereas in 4d models of KSVZ type~\cite{Kim:1979if, Shifman:1979if} we expect these couplings to arise with {\em two} loop factors because they are generated radiatively from gauge field couplings. More detailed discussion of the possible observability of such couplings, including the effects of RG running, can be found in~\cite{Choi:2021kuy}.

\section{Outlook}
\label{sec:outlook}

Axions that arise from gauge fields in extra dimensions offer a compelling solution to the quality problem. Such models have characteristic, qualitative differences from conventional 4d axion models. In particular, they do not have a 4d PQ-breaking phase transition, so their cosmology is of the pre-inflation type. In this paper, I have emphasized that a measurement of the decay constant $f$ for an extra-dimensional axion carries information about the UV cutoff of the theory. A combination of effective field theory estimates~(\S\ref{sec:uvcutoffestimate}); heuristic arguments about geometry, semiclassical solutions, and electric/magnetic duality~(\S\ref{sec:arguments}); and explicit calculations in string theory examples~(\S\ref{sec:examples}) support the idea of a bound on the fundamental UV cutoff of the theory:
\begin{equation}
\Lambda_\textsc{QG} \lesssim 2\cpi \sqrt{S_\mathrm{inst}} f,
\end{equation} 
with $S_\mathrm{inst}$ the instanton action. This is not just an abstract cutoff; it has a specific interpretation as the mass scale of excitations of an axion string, which arises as a fundamental string or a higher dimensional brane wrapping extra-dimensional cycles. For a QCD axion, we have $S_\mathrm{inst} = 8\cpi^2/g^2$ (evaluated at a high scale; while we do not know precisely how $g$ runs above the TeV scale, we can give reasonable estimates), and $f$ can be approximately measured experimentally (the axion mass tells us the value of $f/k_G$, where $k_G$ is an integer expected to be of order $1$). Thus, experimental measurements in axion physics have the potential to tell us about the fundamental quantum gravity cutoff energy in our universe. In quantitative examples in string theory, the fundamental string scale $M_s$ lies no more than about two orders of magnitude above $f$. 

I will close with some brief remarks on interesting directions to pursue in future work:

\begin{itemize}

\item {\bf Wider range of examples.} We have considered several examples of axions associated to moduli with infinite distance limits, all of which have instanton actions and axion string tensions respecting the ``naive'' axion electric and magnetic WGC formulas with $O(1)$ coefficients. We have also considered two examples of axions associated to moduli without associated infinite-distance limits, in the context of Type IIB Calabi-Yau compactifications with a small divisor that can collapse to a point at an edge of moduli space. It would be interesting to explore a wider range of examples, including other singular corners of string theory moduli spaces, examples in Type IIA, examples with warping, and so on. If the expectation of an axion string with tension at the scale~\eqref{eq:TSfsq} continues to hold up across a wide range of examples, this would support our general arguments. On the other hand, if counterexamples are found, it would be interesting to understand their physics in more detail and in particular to see where the heuristic arguments in \S\ref{sec:arguments} break down.

\item {\bf Twists between electric and magnetic states, and axion mixing.}  In \S\ref{subsec:magneticelectric} we gave a heuristic argument for the relationship between tension scales of electrically and magnetically charged objects, but in \S\ref{subsubsec:swisscheese1} we encountered an example where this relationship holds only with a twist: an axion string appeared in the spectrum with the predicted tension, but with nonminimal winding numbers. It would be interesting to understand if there is a general theory behind the nontrivial integer combinations appearing in such examples. More generally, all of the arguments in \S\ref{sec:arguments} could be revisited for the case of multiple axions with mixing.

\item {\bf Interplay with supersymmetry.} The pairing of axions with noncompact scalar moduli (``saxions'') in chiral superfields as in~\eqref{eq:chiralT} has important implications. The scalar moduli can have an important effect on cosmology (see, e.g.,~\cite{Banks:2002sd}). For consistent phenomenology and cosmology, they must be many orders of magnitude heavier than the axion itself. As emphasized in~\cite{Conlon:2006tq, Conlon:2006ur}, this requires that they obtain a mass primarily from K\"ahler potential effects, rather than nonperturbative superpotential terms. This, in turn, implies that supersymmetry breaking effects on Standard Model fields are not sequestered. It would be interesting to explore in greater detail how the axion decay constant, the quantum gravity cutoff, and the scale of supersymmetry breaking are related in various theories.

\item {\bf Ubiquity of axions.} It is often remarked that axions are ubiquitous in string theory, but it is important to sharpen this statement: is the existence of a light axion with $\theta\,\mathrm{tr}(F \wedge F)$ couplings to Standard Model gauge fields a robust prediction of string theory? There are examples in quantum gravity of gauge fields without such couplings, such as Kaluza-Klein gauge fields and the gauge field arising by reducing $C_4$ on the holomorphic 3-cycle of a rigid Calabi-Yau manifold~\cite{Cecotti:2018ufg}. To the best of my knowledge, known examples of 4d gauge fields without an axion coupling do not have light, electrically charged chiral matter fields. I have discussed this briefly in concluding sections of~\cite{Reece:2023czb}. The remarks in \S\ref{subsubsec:monodromyagain} above add support to this picture: we see an explicit clash between a mechanism for adding a large tree-level mass to decouple an axion, and the existence of chiral charged fermion fields in the spectrum of the theory. It would be very interesting to either find a 4d string construction of gauge fields with chiral matter and no axion, or to give an argument from general principles that an axion must exist.\footnote{To be slightly more precise, I mean that the $(-1)$-form instantonic symmetry is gauged, in the language of~\cite{Heidenreich:2020pkc, Aloni:2024jpb}, rather than that an axion exists. This implies either that an axion exists or that there is a chiral symmetry broken only by instanton effects. Along the lines discussed above in \S\ref{subsubsec:greenschwarz}, this can lead to a 4d axion if the chiral symmetry is spontaneously broken. For the Standard Model, a linear combination of the $\SU(2)_\textsc{L}$ and $\Uone_\textsc{Y}$ instantonic symmetries could potentially be gauged by the $\textsc{B}+\textsc{L}$ fermion current, but no such anomalous conserved current exists for $\SU(3)_\textsc{C}$.}

\item {\bf Relation to Swampland conjectures.} The relationship~\eqref{eq:TSfsq} holds whenever both the ``naive'' axion electric and magnetic WGC are satisfied. In general, we expect the WGC to hold with a coefficient that is sensitive to the scalar forces appearing in the theory. For $p$-form gauge theories with $1 \leq p \leq (d-3)$, one can quantify the coefficient by understanding the extremal black holes in the theory, and see that the WGC becomes much stricter when scalar forces are large~\cite{Heidenreich:2015nta}. Alternatively, one can consider the Repulsive Force Conjecture, which requires repulsive gauge forces to overcome all attractive interactions (gravity and scalar) for some objects in the theory~\cite{Arkani-Hamed:2006emk, Palti:2017elp, Heidenreich:2019zkl}. The correct way to quantify the coefficient in the WGC for axions, with $p = 0$ (electric) or $p = (d-2)$ (magnetic), is less clear (but see~\cite{Montero:2015ofa, Bachlechner:2015qja, Brown:2015lia, Heidenreich:2015nta, Hebecker:2018ofv} for proposals involving gravitational instantons in the electric case). My expectation is that there is a form of the axion WGC---both electric and magnetic---that becomes stricter in the presence of saxion interactions much stronger than gravity, and which is (at least parametrically, if not sharply) saturated in all of our examples. It would be interesting to try to formulate such a sharpened conjecture, and to understand its relationship to other conjectured properties of scalar forces in gravitational theories (e.g.,~\cite{Palti:2017elp, Lee:2018spm, Andriot:2020lea, Etheredge:2022opl}). Relatedly, it would be interesting to explore the emergence of kinetic terms for moduli that {\em do not} have infinite-distance limits, like those associated with small cycles in our examples, extending the program initiated in~\cite{Heidenreich:2018kpg, Grimm:2018ohb}, which focused on infinite-distance limits.

\end{itemize}

Axions from extra dimensions offer a strong possibility to connect fundamental physics, including general principles of quantum gravity, with feasible experiments. We should understand them thoroughly to make the most of this opportunity.

\section*{Acknowledgments}

This paper gradually came together over the last few years, and I have likely lost track of some of the conversations that were helpful along the way. Discussions at the 2022 workshop ``Back to the Swamp'' at the IFT in Madrid were very useful. I thank Kiwoon Choi, Jim Cline, Muldrow Etheredge, Naomi Gendler, Jim Halverson, Ben Heidenreich, Luca Martucci, Liam McAllister, Jake McNamara, Jakob Moritz, Mick Nee, Fernando Quevedo, Tom Rudelius, Ben Safdi, and Irene Valenzuela for helpful discussions, questions, or comments at various points along the way. This work was supported in part by the DOE Grant DE-SC0013607.

\appendix

\section{Derivation of 2-cycle and 4-cycle relationship in CY 3-folds}
\label{app:cyderivation}

In this appendix we derive the formula~\eqref{eq:ttaurelation}. We have a set of 2-cycle volumes $t_i$, in terms of which the (tree-level) volume in string units is a cubic function~\eqref{eq:volumettt} and the 4-cycle volumes are derivatives of this function~\eqref{eq:taudef}. From these definitions it immediately follows that (with an implicit sum over $i$) $\tau^i t_i = 3{\cal V}$. We have a K\"ahler potential proportioanl to $k = -2 \log{\cal V}$. Consider the matrix 
\begin{equation}
{\cal M}^{ij} \equiv \frac{\partial \tau^i}{\partial t_j} = n^{ijk} t_k.
\end{equation}
Note that ${\cal M}^{ij} t_j = 2 \tau^i$. Define the inverse matrix ${\cal M}_{ij}$ by ${\cal M}_{ij} {\cal M}^{jk} = \delta^i_k$. We have ${\cal M}_{ij} = \frac{\partial t_i}{\partial \tau^j}$. 
We introduce the notation
\begin{equation}
{\cal K}_{ij} \equiv \frac{\partial^2 k}{\partial \tau^i \partial \tau^j},
\end{equation}
and ${\cal K}^{ij}$ for the inverse matrix, ${\cal K}^{ij}{\cal K}_{jk} = \delta^i_k$. 

Before working out derivatives of $k$ with respect to the 4-cycle volumes, it is easier to first work out derivatives with respect to 2-cycle volumes. Note that, by the relation between $k$ and $\cal V$, we have
\begin{equation}
\frac{\partial}{\partial t_m} k = - \frac{2}{\cal V} \tau^m.
\end{equation}
Hence  we can define a matrix of second derivatives of $k$ with respect to the $t$s:
\begin{equation}
{\widetilde {\cal K}}^{mn} \equiv \frac{\partial^2}{\partial t_m \partial t_n} k = -\frac{2}{\cal V} \frac{\partial \tau^m}{\partial t_n}+ \frac{2}{{\cal V}^2} \tau^m \frac{\partial {\cal V}}{\partial t_n} = -\frac{2}{\cal V} \left({\cal M}^{mn} - \frac{1}{\cal V} \tau^m \tau^n\right).
\end{equation}
Contracting this with  $t_n$, we learn that
\begin{equation}
{\widetilde {\cal K}}^{mn} t_n = \frac{\partial^2 k}{\partial t_m \partial t_n} t_n =  -\frac{2 }{\cal V} \left(2 \tau^m - \frac{1}{\cal V} \tau^m (3 {\cal V})\right) = \frac{2}{\cal V} \tau^m.
\label{eq:goalbutbackwards}
\end{equation}
This has a similar form to what we want to prove.

Next, we claim that the inverse matrix of ${\widetilde{\cal K}}^{mn}$ is
\begin{equation}
{\widetilde{\cal K}}_{mn} = - \frac{\cal V}{2} \left({\cal M}_{mn} - \frac{1}{2{\cal V}} t_m t_n\right).
\end{equation}
This can be checked by direct calculation, but first we need to know that
\begin{equation}
t_i = {\cal M}_{ij} {\cal M}^{jk} t_k = 2 {\cal M}_{ij} \tau^j.
\end{equation}
Using this, we have:
\begin{equation}
 \left({\cal M}_{mn} - \frac{1}{2 {\cal V}} t_m t_n\right)\left({\cal M}^{nl} - \frac{1}{\cal V} \tau^n \tau^l\right) = \delta_m^l - \frac{1}{\cal V} \frac{1}{2} t_m \tau^l - \frac{1}{2{\cal V}} 2 \tau^l t_m + \frac{1}{2{\cal V}^2} t_m \tau^l (3 {\cal V}) = \delta_l^m,
\end{equation}
as claimed.

Acting on both sides of~\eqref{eq:goalbutbackwards} with ${\widetilde {\cal K}}_{lm}$, we obtain
\begin{equation}
t_l = -\left({\cal M}_{lm} - \frac{1}{2{\cal V}} t_l t_m\right) \tau^m.
\end{equation}
This has the form of what we want to prove, {\em if} we can establish that
\begin{equation}\label{eq:toestablish}
{\cal K}_{ij} = \frac{\partial^2 k}{\partial \tau^i  \partial \tau^j} = -\frac{1}{\cal V} \left({\cal M}_{ij} - \frac{1}{2 {\cal V}} t_i t_j\right) = \frac{2}{{\cal V}^2} {\widetilde {\cal K}}_{ij}.
\end{equation}
Next, we evaluate ${\cal K}_{ij}$ by the chain rule. First we have:
\begin{equation}
\frac{\partial k}{\partial \tau^i} = \frac{\partial t_j}{\partial \tau^i} \frac{\partial k}{\partial t_j}.
\end{equation}
Then,
\begin{align}
\frac{\partial^2 k}{\partial \tau^i \partial \tau^j} &= \frac{\partial t_m}{\partial \tau^i} \frac{\partial}{\partial t_m} \left(\frac{\partial t_n}{\partial \tau^j} \frac{\partial k}{\partial t_n}\right) \nonumber \\
&= \frac{\partial t_m}{\partial \tau^i} \frac{\partial t_n}{\partial \tau^j} \frac{\partial^2 k}{\partial t_m \partial t_n} + \frac{\partial t_m}{\partial \tau^i} \frac{\partial k}{\partial t_n} \frac{\partial}{\partial t_m}\left(\frac{\partial t_m}{\partial \tau^j}\right).
\end{align}
We have:
\begin{equation}
\frac{\partial}{\partial t_m}\left(\frac{\partial t_n}{\partial \tau^j}\right) = \frac{\partial}{\partial t_m} {\cal M}_{nj} = - {\cal M}_{np} \left(\frac{\partial}{\partial t_m} {\cal  M}^{pq}\right) {\cal M}_{qj}  = -n^{mpq} {\cal M}_{np} {\cal M}_{qj}.
\end{equation}
So:
\begin{align}
\frac{\partial^2 k}{\partial \tau^i \partial \tau^j} &= {\cal M}_{mi} {\cal M}_{nj} {\widetilde {\cal K}}^{mn} - n^{mpq} \left(-\frac{2}{\cal V}\right) \tau^n {\cal M}_{mi} {\cal M}_{np} {\cal M}_{qj} \nonumber \\
&= -\frac{2}{\cal V} \left[{\cal M}_{ij} - \frac{1}{\cal V} \frac{t_i}{2} \frac{t_j}{2} - n^{mpq} \frac{t_p}{2} {\cal M}_{mi} {\cal M}_{qj} \right] \nonumber \\
&= -\frac{2}{\cal V} \left[{\cal M}_{ij} - \frac{1}{\cal V} \frac{t_i}{2} \frac{t_j}{2} - \frac{1}{2} {\cal M}^{mq} {\cal M}_{mi} {\cal M}_{qj} \right] \nonumber \\
&= -\frac{1}{\cal V} \left({\cal M}_{ij} - \frac{1}{2 {\cal V}} t_i t_j\right).
\end{align}
This establishes~\eqref{eq:toestablish} and completes the proof.

\bibliography{ref}
\bibliographystyle{utphys}

\end{document}